\documentclass[journal]{IEEEtran}
\usepackage[ruled,linesnumbered]{algorithm2e}
\usepackage{cite}
\usepackage{graphicx}
\usepackage{amsmath}
\usepackage{array}
\usepackage{mdwmath}
\usepackage{amssymb}
\usepackage{mdwtab}
\usepackage{stfloats}
\usepackage[footnotesize]{subfigure}
\usepackage{amsmath,amsthm}
\usepackage{threeparttable}
\usepackage{tablefootnote}
\usepackage{color}
\usepackage{url}
\usepackage{algpseudocode}
\usepackage{multicol}
\usepackage{multirow}
\usepackage{setspace}
\usepackage{bm}
\usepackage{enumitem}
\algrenewcommand{\algorithmicrequire}{\textbf{Input:}}
\algrenewcommand{\algorithmicensure}{\textbf{Output:}}
\usepackage{float}

\newtheorem{corollary}{Corollary}

\newtheorem{proposition}{Proposition}

\begin{document}

\renewcommand{\figurename}{Fig.}

\title{Aging Channel Modeling and Transmission Block Size Optimization for Massive MIMO Vehicular Networks in Non-Isotropic Scattering Environment}

\author{Huafu Li,~\IEEEmembership{Graduate~Student~Member,~IEEE,}
	Liqin~Ding,~\IEEEmembership{Member,~IEEE,}
	Yang~Wang,~
	and~Zhenyong~Wang,~\IEEEmembership{Senior~Member,~IEEE}
\thanks{This work was supported in part by the European Union’s Horizon 2020 research and innovation programme under the Marie Skłodowska Curie grant agreement No 887732 (H2020-MSCA-IF VoiiComm), 
in part by the Science and Technology Project of Shenzhen under Grant JCYJ20200109113424990, and in part by the Marine Economy Development Project of Guangdong Province under Grant GDNRC [2020]014.   \emph{(Corresponding author: Yang Wang.)}}
\thanks{Huafu Li and Zhenyong Wang are with the School of Electronics and Information Engineering, Harbin Institute of Technology, Harbin 150001, China (e-mail: \{fairme,  zywang\}@hit.edu.cn).}
\thanks{Liqin Ding is with the Department of Electrical Engineering, Chalmers University of Technology, 412 96 Gothenburg, Sweden (e-mail: liqind@chalmers.se).}
\thanks{Yang Wang is with the School of Electronics and Information Engineering, Harbin Institute of Technology Shenzhen, Shenzhen 518055, China (e-mail: yangw@hit.edu.cn).}}

\maketitle

\begin{abstract}
We investigate the effect of channel aging on multi-cell massive multiple-input multiple-output (MIMO) vehicular networks in a generic non-isotropic scattering environment. Based on the single cluster scattering assumption and the von Mises distribution assumptions of the scatterers' angles, an aging channel model is established to capture the joint effect of spatial and temporal correlations resulting from different angular spread conditions in various application scenarios. Expressions of the user uplink transmission spectral efficiency (SE) are derived for maximum ratio (MR) and minimum mean square error (MMSE) combining. Through numerical studies, the area spectral efficiency (ASE) performance of the network is evaluated in freeway and urban Manhattan road grid scenarios, and easy-to-use empirical models for the optimal transmission block size for ASE maximization are obtained for the evaluated scenarios. 
\end{abstract}


\section{Introduction}

Massive multiple-input multiple-output (MIMO) 
is becoming a reality, but there are still many practical issues to be addressed \cite{sanguinetti2019toward, 6GMobility}. One of them is to understand the impact of channel aging on massive MIMO systems in highly dynamic scenarios and to find corresponding solutions to accommodate such impact for efficient network operation. 
Channel aging refers to the mismatch between the estimated channel coefficients during the channel estimation phase and the actual ones over which the data is transmitted, as wireless channels change inevitably with time \cite{truong2013effects}. Since most of the claimed advantages of massive MIMO rely heavily on the availability of accurate channel state information (CSI) at the base stations (BSs), for precoding in downlink (DL) and combining in uplink (UL) the signals sent / received through the antenna arrays \cite{8094949Capacity, 9113273Prospective}, massive MIMO is more suspensible to channel aging than conventional MIMO employing small arrays. 
For example, it has been shown that a CSI outdated by four milliseconds could cause up to $50$\% degradation in data rate for users with moderate mobility ($30$~km/h)  compared to low mobility ($3$~km/h) when the BS employ arrays with $32$ and $64$ antennas \cite{yin2020addressing}. Therefore, the study of channel aging effects is crucial, especially for applications with highly dynamic environments (e.g., urban scenarios \cite{li2021impact, ding2019kinematic}) or users with high mobility (e.g., ground vehicles \cite{ge2020high} and drones \cite{8214963Drone, 9374639UAV}).  

In the seminal paper \cite{truong2013effects} by K.~T.~Truong \emph{et al.}, an aging channel model  that considers both the channel estimation error and the aging drifts is developed for massive MIMO systems, and a performance analysis framework that covers both UL and DL transmissions is established. The temporal autocorrelation of the channel is assessed based on isotropic scattering (i.e., the Jakes-Clarke model) and equal Doppler shift assumption, resulting in an autocorrelation function (ACF) given by the zeroth-order Bessel function of the first kind. 
Based on this model, the effects of channel aging are then more thoroughly studied by A.~K.~Papazafeiropoulos \emph{et al.} in a series of works \cite{2015Deterministic, kong2015sum, papazafeiropoulos2016impact, 7519076, papazafeiropoulos2017toward}, considering different precoding / combining methods and practical issues such as pilot contamination, phase noise, and hardware impairment. Lately, the study of aging effects has also been extended to non-central network architectures, under the name of distributed antenna system or cell-free massive MIMO network \cite{jiang2021impact, elhoushy2020performance, elhoushy2021limiting, chopra2021uplink, zheng2021impact}. 

To alleviate the effects of channel aging, channel prediction techniques are also proposed, such as the Wiener predictor \cite{truong2013effects, 2015Deterministic, kong2015sum}, Kalman predictor \cite{chopra2021data, kashyap2017performance}, and the autoregressive moving average (ARMA) predictor \cite{younas2020study}. 
Another key measure against channel aging is the optimal design of the channel training frequency, or equivalently, the duration of the transmission block, to ensure good system-level throughput / spectral efficiency (SE) \cite{chopra2016throughput, chopra2017performance, xin2020spectral,2022frameLength}. This problem stems directly from the reasoning that more frequent channel estimation ensures more accurate CSI and that there will be a sweet spot where the resulting performance gain most outweighs the cost. Obviously, such a sweet spot depends on how quickly the channel ages and how much the performance metric of interest is affected by the outdated CSI. 

Both performance evaluation and system design optimization require an aging channel model that correctly captures the spatial and temporal correlations of channel coefficients. However, in most existing works, the assumptions are oversimplified. First, most works adopt the Jakes-Clarke model when assessing the temporal evolution (aging) of the channel, while in practice, the angular spread of multipath components of the channel is typically limited \cite{abdi2002parametric, 3GPP25996, WinnerII}. In other words,  non-isotropic scattering propagation environment is the common case. Measurement campaigns have demonstrated this for BSs in a wide range of scenarios \cite{zhu2018spatial, 3GPP25996} and also for mobile terminals \cite{abdi2002parametric, cheng2013improved} in many scenarios such as the street canyon environment \cite[Section 7.4]{molisch2012wireless}. Our preliminary work \cite{li2021impact} shows that the Jakes-Clarke model may lead to overly pessimistic performance predictions and more than necessary channel training. 
Second, the joint impact of spatial correlation is poorly captured in the existing study. Most works \cite{2015Deterministic,kong2015sum,papazafeiropoulos2016impact,7519076,papazafeiropoulos2017toward, chopra2021data, kashyap2017performance, younas2020study, chopra2016throughput, xin2020spectral,chopra2017performance,elhoushy2020performance, elhoushy2021limiting,jiang2021impact,chopra2021uplink} assume that channels are spatially uncorrelated, and only a few have included spatial correlation in channel modeling. 
For example, when computing spatial correlation matrices, the Laplace distribution and the Gaussian distribution are assumed for the multipath angles in \cite{truong2013effects} and \cite{zheng2021impact}, respectively. In our preliminary work \cite{li2021impact}, spatial correlation is modeled by assuming a uniform distribution within a limited range. However, as we shall discuss in detail later in this paper, the impact of spatial correlation on massive MIMO links is quite complex and requires careful study.

In this paper, we investigate the effect of channel aging on a multi-cell massive MIMO system with vehicle users (VUEs) in a more realistic non-isotropic scattering scenario with spatially correlated channels. We also address the optimal transmission block design problem in typical scenarios such that the area spectral efficiency (ASE) is maximized. Our study focuses on UL transmission and the main contributions are summarized as follows.

\begin{itemize}	
	\item We derive the spatial-temporal cross-correlation (STCC) function of the channel based on the assumption of the von Mises distribution for both angle-of-departure (AoD) and angle-of-arrival (AoA) to represent a general nonisotropic scattering condition, and develop an aging channel model that allows us to study the joint effect of space-time correlation in various application scenarios. The channel model captures the impact of the spatial distribution of BS, VUE, and scatterers with parameters including the central direction and degree of spread of AoA / AoD, as well as the orientation of the BS antenna array and the velocity of the VUE. 

	\item Based on the developed aging channel model, expressions for the VUE's SE performance are derived for both maximum ratio (MR) and minimum mean square error (MMSE) combining, taking into account the channel training overhead and pilot contamination effects.  The system-level ASE performance of the massive MIMO system is then evaluated in freeway and urban Manhattan road grid scenarios. 
	 
	\item Based on numerical studies, we obtain easy-to-use empirical models for the transmission block size for optimizing ASE performance, which turns out to be linear equations of VUE moving speed and square roots of AoD and AoA spread. The performance gain brought by the optimal transmission block design is demonstrated. 
\end{itemize}

The paper organization is as follows. The aging channel model is developed in Section~\ref{sec:system_model}.  Expressions for the UL SE are derived in Section~\ref{sec:3}. The numerical studies in the two scenarios are conducted in Section~\ref{sec:simulation}, where empirical models of the optimal transmission block size are also obtained and evaluated. Finally, Section~\ref{sec:conclusion} concludes this work.

\textbf{Notation:} We use italic lowercase letters, boldface lowercase letters, boldface uppercase letters, and calligraphic letters to represent scalars, column vectors, matrices, and sets, respectively. The expectation operator, the absolute value, the Euclidean norm and the trace operator are denoted by  $\mathbb{E}\left\{  \cdot  \right\}$,  $\left|  \cdot  \right|$, $\left\|  \cdot  \right\|$, and ${\operatorname{tr}}\left(  \cdot  \right)$, respectively.  The conjugate, conjugate transpose, and pseudoinverse operations are denoted by $(\cdot)^{*}$, $(\cdot)^{\mathrm{H}}$, and $(\cdot)^\dag$, respectively. ${{\mathbf{I}}_m}$ stands for the $m\times m$ identity matrix, $j=\sqrt{-1}$, $c$ denotes the speed of light, and a definition is denoted by $\triangleq$. 
Finally, ${\mathcal{N}_\mathbb{C}}({\mathbf{0}},{\mathbf{R}})$ stands for the multi-variate circularly symmetric complex Gaussian distribution with zero mean and covariance matrix ${\bf{R}}$.

\section{Aging Channel Modeling}
\label{sec:system_model}

\begin{figure}[!t]
    \centering
    \includegraphics[width=\linewidth]{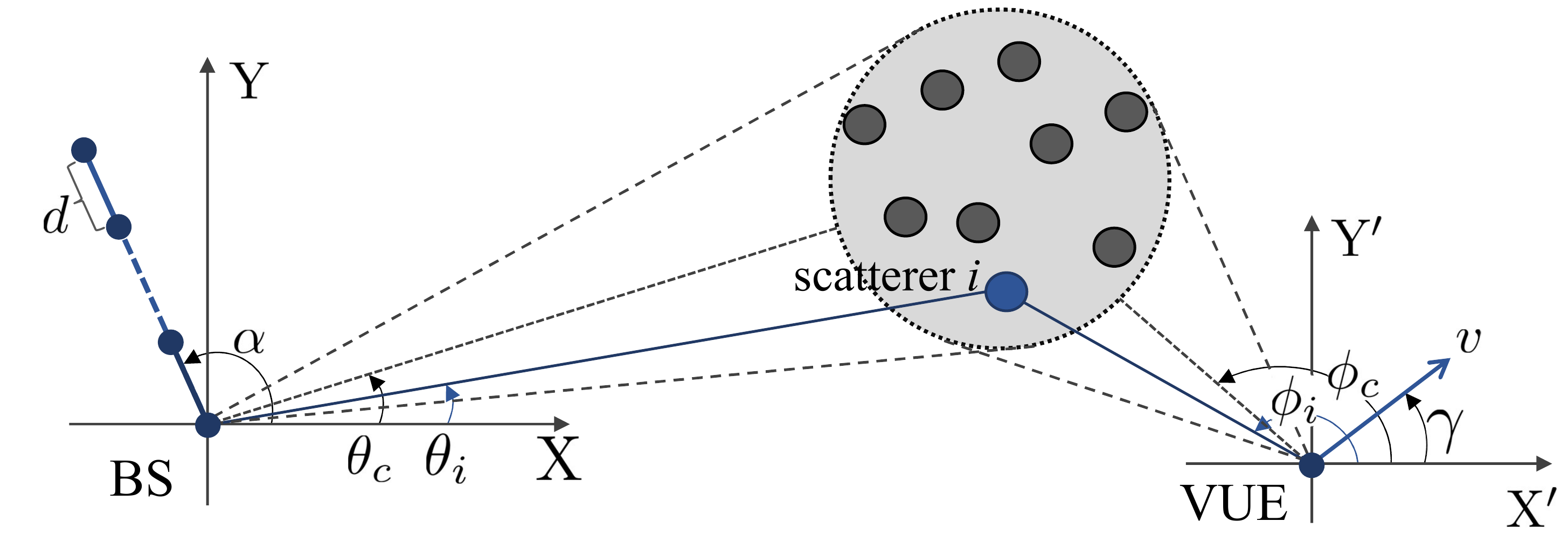}
    \caption{Propagation geometry of a joint space-time correlated channel with one cluster of scatterers.}
    \label{S_T_channelmodel}
\end{figure}

We consider the uplink transmission over one subcarrier in a massive MIMO system. Time is divided into transmission blocks. The pilot sequence is transmitted at the beginning of the transmission block, followed by user data in the rest of the block. We are interested in the impact of channel aging on the performance of the transmission in one block. In this section, we model the narrow-band channel between a single-antenna VUE and its serving BS, where a uniform linear array (ULA) of $M$ antennas is deployed. 

\subsection{Space-Time Cross Correlation Matrix} 
\label{channelmodel}  

As shown in Fig. \ref{S_T_channelmodel}, the orientation of the ULA is specified by an angle $\alpha$ from the arbitrarily selected reference direction $\mathrm{X}$, while the direction of movement of the VUE (of speed $v$) is given by the angle $\gamma$. 
The channel between them is composed of single-bounce multipath elements, caused by a cluster of $S$ static scatterers. The angle of the $i$th scatterer seen from the VUE / BS is denoted by $\phi_{i}$ / $\theta_i$. Since we consider uplink transmission in this paper, $\phi_{i}$ and $\theta_i$ will be referred to as AoD and AoA, respectively. The scatterers are assumed to be distant enough from the BS such that the latter experiences planar wavefronts. Denote the ULA antenna spacing by $d$ and the carrier frequency by $f_c$, and let $\lambda_c = c/f_c$ be the wavelength. 
The $M\times1$ complex channel vector at the time $t$ is therefore given by
\begin{equation}   \label{h_kpt}
\mathbf{h}(t) =  \sum_{i=1}^{S} \xi_{i} \exp \bigg(j 2\pi f_{i}^{D} t + j \frac{2 \pi}{\lambda_{c}} \boldsymbol{\beta}_i  + j\psi_{i}  \bigg), 
\end{equation}
where $\xi_{i}$ is the path gain, $f_{i}^D= {f_c v}\cos (\phi_{i}-\gamma)/{c}$ is the Doppler frequency, $\boldsymbol{\beta}_i$ is the vector of propagation distance difference, whose $p$-th element is given by $(p-1) d \cos \left( {\alpha - {\theta _i}} \right)$, $p = 1,\ldots,M$, and $\psi_{i} \in [-\pi, \pi)$ is the random phase shift associated with the $i$th multipath component. 

We assume that the AoDs $\{\phi_{i}\}$, AoAs $\{\theta_{i}\}$, and phase shifts $\{\psi_{i}\}$ are all independent random variables.
The $M\times M$ STCC matrix of $\mathbf{h}(t)$ with delay $\tau$ is defined by 
\begin{align}
    \mathbf{R}(\tau) \triangleq \frac{ \mathbb{E}\left\{\mathbf{h}(t) \mathbf{h}^{\mathrm{H}}(t+\tau)\right\}}{ \sqrt{ \mathbb{E}\{ \|\mathbf{h}(t)\|^2 \} } \sqrt{ \mathbb{E}\{ \|\mathbf{h}(t + \tau)\|^2 \} } }, 
\end{align} 
and its $(p,q)$-th element can be derived as follows: 
\begin{align} \label{von_S_T}
\left[\mathbf{R}(\tau)\right]_{p, q}= 
\mathbb{E}\left\{ 
\sum_{i=1}^{S} \exp \big[ j a \cos \left(\phi_{i} \!-\!\gamma\right) 
+j b \cos \left(\alpha\!-\!\theta_{i}\right)\big]\right\},
\end{align} 
where 
\begin{equation}
   a = - 2 \pi \tau f_c v/c, \quad b = 2\pi (p - q) d/\lambda_{c}.
\end{equation}
Note that the gains of the multipath components are assumed to be fixed over the time interval of interest. 

In this work, we consider that the number of scatterers approaches infinity ($S\rightarrow \infty$) and that AoD and AoA follow two independent continuous distributions. (In this case, the channel gain of each multipath component is infinitesimal.) In particular, we adopt the empirically verified von Mises distribution for both angles to represent a non-isotropic scattering condition \cite[Section  2.1.2]{gordon2017principles}. 
The PDFs of AoD $\phi \in[-\pi, \pi)$ and AoA $\theta \in[-\pi, \pi)$ are given as follows: 
\begin{align}
    {p}(\phi ) &= \frac{{\exp \left[ {{\kappa_{T} }\cos \left( {\phi  - \phi_c} \right)} \right]}}{{2\pi {I_0}\left( {{\kappa_T}} \right)}}, \\
    {p}(\theta ) &= \frac{{\exp \left[ {{\kappa_R }\cos \left( {\theta  - \theta_c} \right)} \right]}}{{2\pi {I_0}\left( {{\kappa_R}} \right)}}, 
\end{align}
where ${I_0}(z) \triangleq \frac{1}{2\pi} \int_{-\pi}^{\pi} \exp (z \sin x) dx$ is the modified Bessel function of the first kind and zero order \cite[Eq. (9.6.16)]{abramowitz1964handbook}. Following these distributions, the AoDs and AoAs are concentrated around the central directions ${\phi_c} \in [-\pi ,\pi )$ and ${\theta_c} \in [-\pi ,\pi )$, and the degrees of concentration are determined by $\kappa_T$ ($\geq 0$) and $\kappa_R$ ($\geq 0$), respectively. The larger $\kappa_T$ / $\kappa_R$ is, the more concentrated the distribution of AoD / AoA. If $\kappa_T$ / $\kappa_R$ is close to $0$, the distribution is close to uniform. We further adopt 
\begin{equation}
    \sigma_T = \frac{1}{{\sqrt {{\kappa_T}} }}, \quad \sigma_R = \frac{1}{{\sqrt {{\kappa_R}} }}
    \label{sigma2ka}
\end{equation}
as the measure of AoD and AoA spread (in radians) \cite{abdi2002parametric}, since $1/\kappa$ is analogous to the variance in the normal distribution\footnote{In fact, when $\kappa$ is very large, the von Mises distribution approximates closely the normal distribution of variance $1/\kappa$.}.

Based on the above, the $(p,q)$-th element of the STCC matrix $\mathbf{R}(\tau)$, given by \eqref{von_S_T}, can be further derived as the following closed-form\footnote{The relation  $\int_{-\pi}^{\pi} \exp (x \sin z+y \cos z) d z=2 \pi I_{0}(\sqrt{x^{2}+y^{2}} )$  \cite[Eq. (3.338-4)]{Table_of_Integrals2014} is adopted in the derivation.}
\begin{align} 
\left[\mathbf{R}(\tau)\right]_{p, q} 
 = & \int_{-\pi}^{\pi} \exp \{ j a \cos \left(\phi\!-\!\gamma\right) \} 
p(\phi)  \mathrm{d} \phi  \cdot \nonumber\\ 
& \int_{-\pi}^{\pi} \exp \{  j b \cos \left(\alpha\!-\!\theta\right)\}  p(\theta)  \mathrm{d} \theta  \nonumber\\  
= & \rho(\tau) \cdot s(p,q), \label{TCC_All}
\end{align}
where 
\begin{align}\label{ACF_SCF}
    \begin{cases}
    \rho(\tau) = \frac{I_{0}\left(\sqrt{-a^{2}+\kappa_{T}^{2}+2 a j \kappa_{T} \cos \left(\gamma-\phi_{c}\right)}\right)}{I_{0}\left(\kappa_{T}\right)},\\
    s(p,q) = \frac{I_{0}\left(\sqrt{-b^{2}+\kappa_{R}^{2}+2 b j \kappa_{R} \cos \left(\alpha-\theta_{c}\right)}\right)}{I_{0}\left(\kappa_{R}\right)}.
    \end{cases}
\end{align}
It can be easily verified that $\rho(\tau)$ is the ACF of any entry of the time-varying channel $\mathbf{h}(t)$, and that $s(p,q)$ is the spatial correlation function (SCF) between the $p$-th and $q$-th entries at any time instance. 
We can see from \eqref{ACF_SCF} that $\rho(\tau)$ ($0\leq |\rho(\tau)| \leq 1$) is affected by the AoD spread, the speed of the VUE and its moving direction relative to the AoA central direction (i.e., $\gamma -\phi_c$), besides the time delay; and that $s(p,q)$ ($0\leq |s(p,q)| \leq 1$) is affected by the AoA spread, the spacing of the ULA at the BS and its orientation relative to the AoD central direction (i.e., $\alpha -\theta_c$). Together, these parameters determine the STCC matrix. 

This analytically tractable model allows us to investigate the joint effects of the spatial-temporal characteristics of channels on massive MIMO networks under a wide range of scattering conditions by adjusting the above mentioned parameters, and hopefully, draw some far-reaching conclusions.  
For example, in the case of $\kappa_T =0$, $\rho(\tau) = {J_0}\left( 2 \pi \tau f_c v/c \right)$ is immediately obtained, where $J_0(z) \triangleq \frac{1}{2\pi} \int_{-\pi}^{\pi} \exp (-j z \sin x) d x$ is zeroth-order Bessel function of the first kind \cite[Eq. (9.19)]{temme1996special}, which is the Jakes-Clarke ACF model based on the isotropic scattering assumption ($p(\phi) = 1/2\pi$) around the mobile user. 
In the case of $\kappa_R =0$ ($p(\theta) = 1/2\pi$), the SCF becomes $s(p,q) = J_{0}\left[ 2 \pi (p - q) d / \lambda_{c} \right]$, from which it can be seen that although the BS is surrounded by a large number of isotropic scatterers, there can still be correlation among the antennas. This is due to the fact that ULAs have better angular resolution for the boresight direction than for those directions closer to the end-fire \cite{bjornson2017massive}. 

Finally, we remark that it is reasonable to assume that the scatterers that contribute to the multipath components of the channel are unchanged within the time interval of our interest (i.e.,, the duration of a transmission block), which typically lasts several milliseconds \cite{xia2020learning, marzetta2016fundamentals}. With this assumption, the STCC matrix $\mathbf{R}(\tau)$ with elements given by \eqref{TCC_All} is in fact time independent. 
We also note that the total gain of the channel, $G(t) \triangleq \|\mathbf{h}(t)\|^2$, is also considered to be unchanged in this short time interval and is below denoted by $G$.

\subsection{Numerical Examples}
\label{sec:2b}

\begin{figure}[!t]
    \centering
    \includegraphics[width=3.45in]{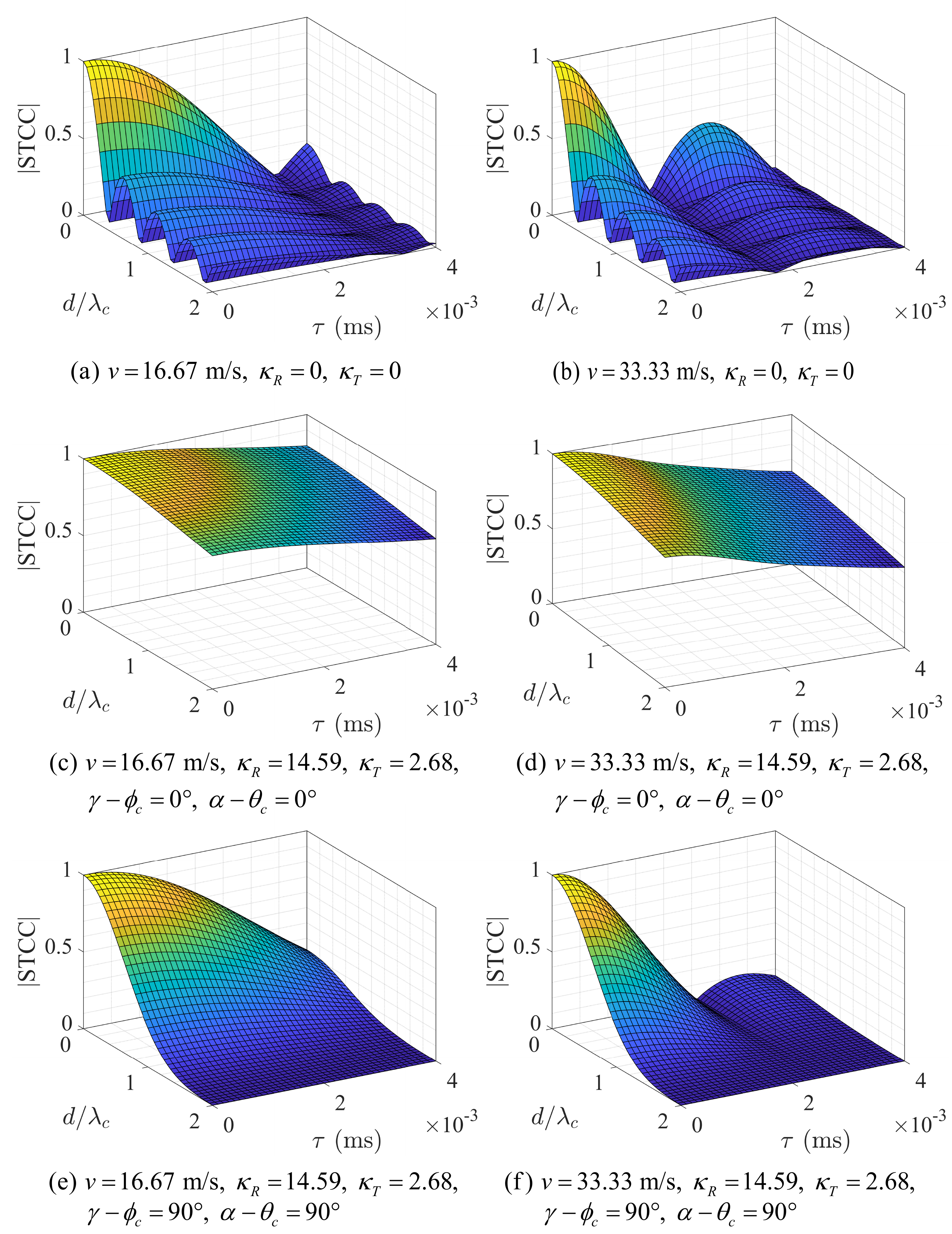}
    \caption{Numerical results of space-time correlation for different scattering and mobility conditions.   }
    \label{S_T}
\end{figure}

In Fig.~\ref{S_T}, the absolute value of the STCC for two adjacent antenna elements ($p-q=1$), as a function of antenna spacing $d$ and delay $\tau$, is plotted under different conditions resulting from the combination of the following parameters: 
\begin{itemize}
    \item Isotropic scattering: $\kappa_T =0$, $\kappa_R =0$, and non-isotropic scattering: $\kappa_T =2.68$, $\kappa_R =14.59$ (corresponding to $\sigma_{T}\approx35^{\circ}$ and $\sigma_{R}\approx15^{\circ}$, as suggested in \cite[Table 5-4]{WinnerII} for the urban microcell NLOS scenario);
    \item Different relative orientation of the ULA at BS: $\alpha-\theta_c=0^{\circ}$ or $90^{\circ}$, and relative direction of motion of the VUE: $\gamma-\phi_c= 0^{\circ} $ or $90^{\circ}$;
    \item Moderate speed $v = 16.67$ m/s ($60$ km/h), and high speed $v=33.33$ m/s ($120$ km/h).
\end{itemize}
Note that when $\kappa_T=\kappa_R=0$, the results are not affected by $\gamma$, $\phi_{c}$, $\alpha$, and $\theta_{c}$.

By comparing Fig.~\ref{S_T} (a) and (b) with rest of the subplots, it can be clearly seen that the isotropic scattering assumption leads to the fastest decrease in the ACF and SCF, and in the considered non-isotropic scattering environment, the fluctuation of the STCC is much smoother. In fact, the larger $\kappa_T$ is, the stronger the time auto-correlation will be. In this case, the channel ages slowly. Similarly, the larger $\kappa_R$ is, the stronger the spatial correlation will be. However, a large SCF is generally considered to hinder the performance of a massive MIMO system \cite{zheng2021impact} while a large ACF can better withstand channel aging impairments \cite{li2021impact}. 
Furthermore, a faster moving speed $v$ produces a smaller temporal correlation, which leads to a stronger aging effect, as can be observed by comparing the subplots in the two columns. 
In addition, by comparing Fig.~\ref{S_T}(c) and (d) with (e) and (f), it is found that the relative orientation of the BS ULA and the relative moving direction of the VUE also substantially affect SCF and ACF. In particular, when $\gamma-\phi_c=90^{\circ}$ / $\alpha-\theta_c=90^{\circ}$, a much smaller ACF / SCF is obtained, which can be inferred from \eqref{ACF_SCF} directly. 
Therefore, it should remembered that under non-isotropic scattering conditions there are more parameters affecting the STCC in addition to the AoA / AoD spread.

\subsection{The Discrete Aging Channel Model} 

Since the transmission block is further divided into symbols, a discretized channel model is required for further discussion. In particular, we assume that a transmission block contains $C$ symbols in total and that the length of a symbol is denoted by $T_s$. Moreover, we let $\mathbf{h}[n] \triangleq \mathbf{h}(nT_s)$, $n =1,\ldots, C$. 

The assumptions in Section \ref{channelmodel} implies that the entries of $\mathbf{h}(t)$ can be modeled as zero-mean complex Gaussian processes. Since $\mathbb{E}\left\{ {{\mathbf{h}(t)}{{\mathbf{h}^\mathrm{H}(t)}}} \right\} = G \cdot \mathbf{R}_0$ where $\mathbf{R}_0 \triangleq \mathbf{R}(\tau =0)$ is the positive semidefinite spatial correlation matrix (and also the covariance matrix due to the zero mean), indicating that we can model the channel as $\mathbf{h}[n] \sim  \mathcal{N}_\mathbb{C}(\mathbf{0}, G \cdot \mathbf{R}_0 )$. Moreover, \eqref{TCC_All} suggests that the time and spatial correlation can be decoupled. Therefore, following \cite{chopra2017performance}, we model the channel vector at the $n$-th symbol time as a linear combination of $\mathbf{h}[0]$, initial state of the channel, and an independent innovation component $ {{\mathbf{z}}}[n]$ that follows the same distribution, i.e., ${{\mathbf{z}}}[n] \sim  \mathcal{N}_\mathbb{C}(\mathbf{0}, G \cdot \mathbf{R}_0 )$: 
\begin{equation}
	{{\mathbf{h}}}[n] = {\rho}[n] {{\mathbf{h}}}[0] + {\bar \rho}[n]{{\mathbf{z}}}[n]. 
	\label{channelaingbasic}
\end{equation}
Moreover, $\mathbf{h}[0]$ can be generated by 
\begin{equation}
\mathbf{h}[0]= (G \cdot \mathbf{R}_0 )^{1 / 2} \mathbf{h}^\mathrm{w}, 
\end{equation}
where $\mathbf{h}^\mathrm{w} \sim  \mathcal{N}_{\mathbb{C}}\left(\mathbf{0}, \mathbf{I}_{M}\right)$ is an $M\times1$ circularly symmetric white complex Gaussian vector, $\rho[n] \triangleq \rho(\tau = nT_s)$, and 
${\bar \rho}[n] \triangleq \sqrt {1 - |\rho[n]|^2}$,

\section{Uplink Transmission Performance Analysis}
\label{sec:3}

We consider a multi-cell massive MIMO network which consists of $L$ BSs and operates according to a synchronous time-division duplex (TDD) protocol. Identical ULAs are deployed at all BSs. $K$ single-antenna VUEs transmit in UL on a shared frequency subcarrier. Each VUE is associated to a serving BS following certain rules, e.g., based on the received signal strength, while a BS serves multiple VUEs. We assume that VUE $k$ is served by BS $l_k$ and focus on receiving at this particular BS in this section. We use the subscript $(\cdot)_k$ to specify the channel and the related parameters associated with VUE $k$. As already mentioned, time is divided into transmission blocks and each transmission block lasts $C$ symbols. In particular, we assume that the first $T$ symbols are used for the transmission of the pilot sequence, and the subsequent $C-T$ symbols are used for the transmission of data. Moreover, $T \ll (C-T)$ is assumed.

\subsection{Uplink Channel Training}
\label{uplinktranining}

$T$ mutually orthogonal pilot sequences $\{ \boldsymbol{\varphi}_{1}, \ldots ,\boldsymbol{\varphi}_{T} \in \mathbb{C}^T\}$, with ${\left\| {{\boldsymbol{\varphi}_{t}}} \right\|^2} = T$ for $t=1,\ldots,T$, are adopted for channel training \cite{sanguinetti2019toward}. They are randomly assigned to the VUEs and transmitted by the VUEs to BS $l_k$ simultaneously during the channel training phase. 
We assume that $K > T$, which means that some VUEs are assigned the same pilot and channel training suffers from pilot contamination. 
Let $\boldsymbol{\varphi}_{t_k}$ be the pilot assigned to VUE $k$, and $\mathcal K_{t_k}$ as the index set of the VUEs sharing this pilot. 
Since $T$ is usually a very small number,  the aging issue is not considered for the training phase\footnote{It is shown in \cite{7519076} that the additional error introduced into the channel estimation by the aging effect in the training stage is negligible. The same assumption is adopted in  \cite{truong2013effects,papazafeiropoulos2016impact,jiang2021impact}.}.
Namely, $\mathbf{h}[n] \approx \mathbf{h}[0]$ for $n = 1,\ldots, T$ is assumed.

The received signal at the BS during the training phase over $T$ consecutive symbols, $\mathbf{Y}^{\mathrm{p}} \in \mathbb{C}^{M \times T}$, is given by 
\begin{equation}
    {\mathbf{Y}}^{{\mathrm{p}}} =  \sum\limits_{k' \in \mathcal{K}} \sqrt{P_{k'}}  \, {{\mathbf{h}}_{k'}[0]} \boldsymbol{\varphi}_{{t_{k'}}}^{\mathrm{T}} + {{\mathbf{W}}^{{\mathrm{p}}}},
\end{equation}
where $\mathcal{K} \triangleq \{1,\ldots, K\}$, ${P_{k'}} \ (\geqslant0)$ is the transmit power of VUE ${k'}$, and ${\mathbf{W}}^{{\mathrm{p}}}  \in {\mathbb{C}^{M \times T}}$ is the noise matrix with i.i.d. elements following $\mathcal{N}_{\mathbb{C}}\left(0, \sigma_n^{2}\right)$. 

To estimate the channel for VUE $k$, i.e., $ {{\mathbf{h}}_{k}[0]}$, the BS first cancels the signal from other VUEs that are assigned with orthogonal pilots by multiplying ${\mathbf{Y}}^{{\mathrm{p}}}$ using the normalized conjugate of $\boldsymbol{\varphi}_{t_k}$. By doing this, the following processed received signal  ${\mathbf{y}}_k^{{\mathrm{p}}} \in \mathbb{C}^{M\times1}$ is obtained: 
\begin{align}
  \mathbf{y}_{k}^{\mathrm{p}}  = \frac{1}{\sqrt{T}} \mathbf{Y}^{\mathrm{p}} \boldsymbol{\varphi}_{t_k}^{*} 
  & =  \sum\limits_{k' \in \mathcal{K}_{t_k}} {\sqrt{P_{k'} T} }\, {{\mathbf{h}}_{k'}}[0] + {{\mathbf{n}}^{\mathrm{p}}_{k}},  \label{processedreceivedpilotsignal}
\end{align}
where $\mathbf{n}^{\mathrm{p}}_{k} =  \frac{1}{\sqrt{T}} \mathbf{W}^{\mathrm{p}} \boldsymbol{\varphi}_{t_k}^{*}  \sim \mathcal{N}_{\mathbb{C}}\left(\mathbf{0}, \sigma_n^{2} \mathbf{I}_{M}\right)$. 
Using the MMSE estimation \cite{van2018large,Channel_Estimation6415397}, an estimate of $\mathbf{h}_{k}[0]$ is obtained:
\begin{equation}
    \widehat{\mathbf{h}}_{k}[0] = \sqrt {P_k T} {G_k \mathbf{R}}_{0,k} {\mathbf{\Psi }}_k^{ - 1} {\mathbf{y}}_k^{{\mathrm{p}}} \sim \mathcal{N}_{\mathbb{C}}\left(\mathbf{0}, \mathbf{\Phi}_k\right), 
    \label{estimatedchannel}
\end{equation}
where  
\begin{align}\label{eq:Psi_k}
\boldsymbol{\Psi}_k = \mathbb{E} \left\{\mathbf{y}_k^{\mathrm{p}} ( \mathbf{y}_k^{\mathrm{p}} )^{\mathrm{H}} \right\} = \sum_{k' \in \mathcal{K}_{t_k}} P_{k'} T G_{k'}\mathbf{R}_{0,k'}+ \sigma_n^{2} \mathbf{I}_{M}, 
\end{align}
and 
\begin{align}\label{eq:Phi_k}
    {{\mathbf{\Phi}}_{k}} ={P_k}T{G_{k}^{2}\mathbf{R}}_{0,k} {\mathbf{\Psi }}_k^{- 1}{\mathbf{R}}_{0,k}. 
\end{align} 
The estimation error 
\begin{align} \label{eq:Est_error}
    {\widetilde {\mathbf{h}}_{k}}[0] \triangleq {{\mathbf{h}}_{k}}[0]  - {\widehat {\mathbf{h}}_{k}}[0] \sim \mathcal{N}_{\mathbb{C}}\left(\mathbf{0},G_{k} \mathbf{R}_{0,k}-\boldsymbol{\Phi}_{k} \right) 
\end{align}
is independent of ${\widehat {\mathbf{h}}_{k}}[0] $ due to the properties of MMSE estimation.

We remark that if the BSs adopt MMSE combining during the data transmission phase, each of them will need to estimate the channels and spatial correlation matrices for all users in the system, that is, $\{ \widehat{\mathbf{h}}_{k'}[0] \}_{k' \in\mathcal K}$ and $\{ \mathbf{R}_{0,k'} \}_{k' \in\mathcal K}$\footnote{It has been shown that in practice, a small number of extra time-frequency resources is enough to estimate the spatial correlation matrices using, for example, the regularization approach \cite[Sect. 3.3.3]{bjornson2017massive} or the staggered pilot method \cite{upadhya2018covariance}. More estimation methods are discussed in \cite[Remark 2]{bjornson2014massive}.}. This requires that every BS knows which pilot is assigned to VUE $k'$, $\forall k' \in\mathcal K$, and in practice the pilot information can be shared via the X2 interface that connects each BS \cite{7234933X2}.

\subsection{Uplink Data Transmission}
\label{uplinkdatatransmission}

During the subsequent UL data transmission phase,  all VUEs transmit their data again simultaneously, and the signal received at BS $l_k$ is given by
\begin{equation}
    \mathbf{y}^{\mathrm{d}}[n]=\mathbf{h}_{k}[n] x_{k}[n]+\sum_{k' \in \mathcal{K} \setminus\{k\}} \mathbf{h}_{k'}[n] x_{k'}[n]+\mathbf{w}[n],
    \label{eq06}
\end{equation}
for $n = 1,\ldots, C-T$, where ${x_{k'}}\left[ n \right]\sim {\mathcal{N}_\mathbb{C}}\left( {0,{P_{k'}}} \right)$  is the signal transmitted from VUE ${k'}$, and ${{\mathbf{w}}}\left[ n \right] \sim {\mathcal{N}_\mathbb{C}}\left( {{\mathbf{0}},{\sigma_n^2}{{\mathbf{I}}_M}} \right)$ is the noise vector.

Based on \eqref{channelaingbasic} and \eqref{eq:Est_error}, the BS models the actual channel vector for VUE $k$ for the $n$-th data symbol as 
\begin{align}
\mathbf{h}_{k}[n] &=\rho_{k}[n]\left(\widehat{\mathbf{h}}_{k}[0]+\widetilde{\mathbf{h}}_{k}[0]\right)+\bar{\rho}_{k}[n] \mathbf{z}_{k}[n]   \nonumber\\
&=\rho_{k}[n] \widehat{\mathbf{h}}_{k}[0]+\widetilde{\mathbf{e}}_{k}[n],
\label{realchannel}
\end{align}
where $\widetilde{\mathbf{e}}_{k}[n]$ is the difference between the actual channel and the deterministic 
part $\rho_{k}[n] \widehat{\mathbf{h}}_{k}[0]$  and is given by 
\begin{align}
\widetilde{\mathbf{e}}_{k}[n] = \rho_{k}[n] \widetilde{\mathbf{h}}_{k}[0]+\bar{\rho}_{k}[n] \mathbf{z}_{k}[n] \sim {\mathcal{N}_\mathbb{C}} ( {{\mathbf{0}},{{\mathbf{Q}}_{k}} }  )    
\end{align} 
with 
\begin{equation}
    \mathbf{Q}_{k}= G_{k}\mathbf{R}_{0,k}-|\rho_{k}[n]|^{2} \mathbf{\Phi}_{k}.
    \label{E_kl}
\end{equation} 
We call  $\widetilde{\mathbf{e}}_{k}[n]$ the accumulative error (from the channel estimation error and channel aging effect). Note that $\widetilde{\mathbf{e}}_{k}[n]$ is independent of ${\widehat {\mathbf{h}}_{k}}[0]$. 
The BS then selects a receive combining vector ${\mathbf{v}}_{k}[n] \in \mathbb{C}^{M\times1}$ according to \eqref{realchannel} for VUE $k$ and symbol $n$, multiplies it with the received signal $\mathbf{y}^{\mathrm{d}}[n]$ to obtain 
\begin{align}
    \widetilde{x}_{k}[n]   
    = & \mathbf{v}_{k}^{\mathrm{H}}[n] \mathbf{h}_{k}[n] x_{k}[n] + \sum_{k' \in \mathcal{K}\setminus\{k\}} \mathbf{v}_{k }^{\mathrm{H}}[n] \mathbf{h}_{k'}[n] x_{k'}[n] \nonumber \\ 
    &  +\mathbf{v}_{k}^{\mathrm{H}}[n] \mathbf{w}[n],
\label{xhat}
\end{align}
and estimates ${x_k}[n]$ from ${{\widetilde x}_k}[n]$.

\subsection{Discussion on the Effect of AoD/AoA Distribution} 
\label{sect:scf_impact}

The effect of AoD and AoA distributions on the accumulative error $\widetilde{\mathbf{e}}_{k}[n]$ is rather complicated.
First of all, when the channel is assumed to be non-aging,  i.e., with  $\rho_{k}[n]=1$ in \eqref{realchannel}, $\widetilde{\mathbf{e}}_{k}[n]$ contains the channel estimation error only  \cite{marzetta2016fundamentals,bjornson2017massive}. However, as we can see, severe channel aging effects (small $|\rho_{k}[n]|$) can significantly enlarge the variances of the accumulative error. Recall that a smaller AoD spread (corresponding to a larger $\kappa_T$) causes a slower aging rate. In the following discussions, we treat $\rho_{k}[n]$ as given.

We may use the normalized mean-squared error (NMSE), defined as follows 
\begin{align}
\mu_{k}[n] \triangleq \frac{\mathbb{E}\big\{\left\| \widetilde{\mathbf{e}}_{k}[n]  \right\|^{2}\big\}}{\mathbb{E}\big\{\left\|\mathbf{h}_{k}[n]\right\|^{2}\big\}}  
\end{align} 
to quantify the impact of the accumulative error for VUE $k$ at symbol time $n$, and we wish the value of $\mu_{k}[n]$ to be small. 
Letting $\zeta_k \triangleq {P_{k} T G_{k}}/{\sigma_n^{2}}$,  $\boldsymbol{\Theta}_{k}^{\mathrm{npc}} = \zeta_k \mathbf{R}_{0, k}+  \mathbf{I}_{M}$, 
and $\boldsymbol{\Theta}_{k}^{\mathrm{pc}} =  \sum_{k' \in \mathcal{K}_{t_k}\setminus \{k\} } \zeta_{k'} \mathbf{R}_{0,k'}  $  
and using  \eqref{eq:Psi_k}, \eqref{eq:Phi_k}, and \eqref{E_kl}, it can be shown that 
\begin{align}
\mu_{k}[n] 
&=1-\frac{|\rho_{k}[n]|^{2} \zeta_k \operatorname{tr}\left(\mathbf{R}_{0,k} 
\left( \boldsymbol{\Theta}_{k}^{\mathrm{npc}} + \boldsymbol{\Theta}_{k}^{\mathrm{pc}} \right)^{-1} 
\mathbf{R}_{0, k}\right)}{\operatorname{tr}\left(\mathbf{R}_{0,k}\right)} \nonumber \\ 
&\geq 1-\frac{|\rho_{k}[n]|^{2} \zeta_k \operatorname{tr}\left(\mathbf{R}_{0,k} 
\left( \boldsymbol{\Theta}_{k}^{\mathrm{npc}} \right)^{-1}  \mathbf{R}_{0,k}\right)}{\operatorname{tr}\left(\mathbf{R}_{0,k}\right)}, \label{nonPC}
\end{align} 
and the equality in \eqref{nonPC} holds if $\mathcal{K}_{t_k} =  \{k\} $, that is when no other user is sharing the pilot with VUE $k$, or if $\mathbf{R}_{0, k} \mathbf{R}_{0, k'}=\mathbf{0}_{M \times M}$, $\forall k' \in  \mathcal{K}_{t_k}\setminus \{k\}$, namely, when the spatially correlation matrices for the contaminating users are orthogonal to that for VUE $k$. 
Based on the assumptions in Section~\ref{channelmodel}, it can be shown that when the central directions $\theta_c$ of the scattering components are more separated and when the AoA spread is smaller (when $\kappa_R$ is larger) for users belonging to $\mathcal{K}_{t_k}$, the eigenspaces of $\{ \mathbf{R}_{0,k'} \}_{k' \in  \mathcal{K}_{t_k}}$ are less aligned. This means that the impact of pilot contamination on channel estimation would be less severe in this case.

Substituting the eigenvalue decomposition  $\mathbf{R}_{0,k} = \mathbf U \boldsymbol{\Lambda}_{k} \mathbf U^{-1}$, where $\mathbf U$ is formed by the eigenvectors and the diagonal matrix $\boldsymbol{\Lambda}_{k}=\operatorname{diag}\left(\lambda_{k, 1}, \ldots, \lambda_{k, M}\right)$ contains the non-negative eigenvalues in its diagonal elements, \eqref{nonPC} can be further written as
\begin{align} 
\mu_k^{\mathrm{npc}}[n]  
& \triangleq 1-\frac{  |\rho_{k}[n]|^{2} \zeta_k \operatorname{tr}\left(\boldsymbol{\Lambda}_{k}\left(\zeta_k \boldsymbol{\Lambda}_{k}+\mathbf{I}_{M}\right)^{-1} \boldsymbol{\Lambda}_{k}\right)}{M}   \nonumber\\ \label{NMSE_1}
& = 1-\frac{ |\rho_{k}[n]|^{2}  \zeta_k }{M} \sum_{m=1}^{M} \frac{\lambda_{k, m}^{2}}{ \zeta_k \lambda_{k, m}+1}.
\end{align}
Note that $\sum_{m=1}^{M} \boldsymbol{\Lambda}_{k, m} = \operatorname{tr}(\mathbf{R}_{0, k})=M$ always holds. 
Based on \cite[Lemma 4.2]{cellFree}, it can be proved that the  Hessian matrix of $\mu_k^{\mathrm{npc}}[n]$ (as a function of $\lambda_{k,1}, \ldots, \lambda_{k,M}$) is negative definite and strictly concave for all $\lambda_{k, m}$. 
Moreover, the maximum value of  $\mu_k^{\mathrm{npc}}[n]$ is achieved when $\lambda_{k, m}=1$ for all $m$, while the minimum value is achieved when $\mathbf{R}_{0,k}$ is a rank-one matrix, which occurs when $\kappa_R \rightarrow \infty$ (AoA spread is $0$). The channel for UVE $k$ is spatially fully correlated in the latter case. However,  it is well known that a small AoA spread leads to a rank-deficient channel matrix, which reduces the channel capacity.

\subsection{User's SE Performance Analysis}
\label{sec:performance_analysis}

Based on the models developed in the previous section and following the analysis methods adopted in \cite{sanguinetti2019toward,chopra2017performance, zheng2021impact}, the user's achievable SE for the UL transmission over the aging channel is derived.

\begin{proposition}\label{prop1}
In the considered multi-cell multi-user massive MIMO network, the average achievable SE for VUE $k$ in one transmission block over the aging channel \eqref{channelaingbasic} 
is given by 
\begin{equation}
    \mathrm{SE}_{k}=\frac{1}{C} \sum_{n=1}^{C-T} \mathrm{SE}_{k}[n], 
    \label{SE}
\end{equation} 
where $\mathrm{SE}_{k}[n]$ stands for the achievable SE at symbol $n$ and is given by 
\begin{equation}
    \mathrm{SE}_{k}[n] = \log_{2}\left(1+ \eta_{k}[n]\right). 
    \label{Ins_SE}
\end{equation}
The instantaneous effective signal-to-interference-and-noise ratio (SINR) of the $n$-th data symbol, ${\eta _k}[n]$, can be calculated as 
\begin{equation}
    \eta _k[n] = \frac{{{P_k}  {{\Big| {\rho_{k}[n] {\mathbf{v}}_{k}^{\emph{H}}[n]{{\widehat {\mathbf{h}}}_{k}}[0]} \Big|}^2}}}{{\sum\limits_{k' \in \mathcal{K}\setminus \{ k\} } {{P_{k'}}  {{\Big| {\rho _{k'}[n] {\mathbf{v}}_{k}^{\emph{H}}[n]{{\widehat {\mathbf{h}}}_{k'}}[0]} \Big|}^2}}  + {\mathbf{v}}_{k}^{\emph{H}}[n]{{\mathbf{E}}}{{\mathbf{v}}_{k}}[n]}}
     \label{SINRnew}
\end{equation}
where 
\begin{equation}
    {{\mathbf{E}}} = \sum\limits_{k' \in \mathcal{K}} {{P_{k'}}{{\mathbf{Q}}_{k'}} + {\sigma_n^2}{{\mathbf{I}}_M}}.
\end{equation}
and $\mathbf{Q}_{k'}$ is given by \eqref{E_kl}.
\end{proposition}

\begin{IEEEproof}
Substituting \eqref{realchannel} into \eqref{xhat}, we obtain
\begin{equation}
    {\widetilde x_k}[n] = \underbrace {{\rho_{k}}[n]{\mathbf{v}}_{k}^{\mathrm{H}}[n]{{\widehat {\mathbf{h}}}_{k}}[0]{x_k}[n]}_{\text{desired signal}} + \underbrace {\mathrm{IN}_{k}[n]}_{\text{interference plus noise}},
    \label{eq11}
\end{equation}
where the interference-plus-noise term is given by 
    \begin{align} 
    \mathrm{IN}_{k}[n]  = & \sum_{k' \in \mathcal{K} \backslash\{k\}} \rho_{k'}[n] \mathbf{v}_{k}^{\mathrm{H}}[n] \widehat{\mathbf{h}}_{k'}[0] x_{k'}[n] \nonumber \\
    & +\sum_{k' \in \mathcal{K}} \mathbf{v}_{k}^{\mathrm{H}}[n] \widetilde{\mathbf{e}}_{k'}[n] x_{k'}[n]+\mathbf{v}_{k}^{\mathrm{H}}[n] \mathbf{w}[n]. \label{eq:IN}
    \end{align}

The processed signal in \eqref{eq11} can be treated as a discrete memoryless interference channel with random channel response $\rho_k[n]{\mathbf{v}}_{k}^{\mathrm{H}}[n]{\widehat {\mathbf{h}}_{k}}[0]$. 
We note that $\mathrm{IN}_{k}[n]$ has zero mean, since the signals $x_{k'}[n]$ and the noise $\mathbf{w}[n]$ are independent of the realizations of the channel estimates and estimate errors and have zero mean. Hence, the variance of $\mathrm{IN}_{k}[n]$ can be calculated as
\begin{align}
        \mathbb{V} \{\mathrm{IN}_{k}[n] \} = & \sum_{k' \in \mathcal{K} \backslash\{k\}} P_{k'} \Big| \rho_{k'}[n] \mathbf{v}_{k}^{\mathrm{H}}[n] \widehat{\mathbf{h}}_{k'}[0]\Big|^{2} \nonumber\\
         + \sum_{k' \in \mathcal{K}} & P_{i} \mathbf{v}_{k}^{\mathrm{H}}[n] \mathbf{Q}_{k'} \mathbf{v}_{k}[n]+\sigma_n^{2} \mathbf{v}_{k}^{\mathrm{H}}[n] \mathbf{I}_{M} \mathbf{v}_{k}[n].
        \label{IN}
\end{align}
Finally, using \eqref{eq11} and \eqref{IN} yields \eqref{SINRnew}.
\end{IEEEproof}
 
We note that by letting $\rho_{k'}[n]=1 $ for all $k'$ and $n$, equation \eqref{SINRnew} gives us the instantaneous SINR over the non-aging channel, 
as given by \cite[Eq. (41)]{sanguinetti2019toward}, and \eqref{SE} also yields the average achievable UL SE in this setting. 

When the MR combining scheme is adopted by the BS, the following combining vector would be used: 
\begin{equation}
    {\mathbf{v}}_{k}^{\mathrm{MR}}[n]= \mathbf{\widehat h}_{k}[0].
    \label{MR}
\end{equation}
MR combining maximizes the power of the desired signal $x_k[n]$ but is prone to interferences that constitute \eqref{eq:IN}. 
To maximize the instantaneous SINR, BS can instead adopt the MMSE combining.  

\begin{corollary}
The instantaneous SINR given by \eqref{SINRnew} is maximized when the following MMSE combining vector is adopted:
\begin{equation}
    {\mathbf{v}}_{k}^{{\mathrm{MMSE}}}[n] = P_k \Big( {\sum\limits_{k' \in \mathcal{K}} {{P_{k'}}} |\rho_{k'}[n]|^2 {{\widehat {\mathbf{h}}}_{k'}}[0]\widehat {\mathbf{h}}_{k'}^{\mathrm{H}}[0] + {{\mathbf{E}}}} \Big)^\dag \, {\widehat {\mathbf{h}}_{k}}[0].
    \label{V_MMSE}
\end{equation}
\end{corollary}

\begin{IEEEproof}
We note that the SINR in \eqref{SINRnew} has the form of a generalized Rayleigh quotient: $\hbar  = \left|\mathbf{v}^{\mathrm{H}} \mathbf{x}\right|^{2} / \mathbf{v}^{\mathrm{H}} \mathbf{B} \mathbf{v} = \mathbf{x}^{\mathrm{H}} \mathbf{B}^{-1} \mathbf{x}  $, where   $\mathbf{x} \in \mathbb{C}^{M}$ is an arbitrary nonzero vector and  
$\mathbf{B} \in \mathbb{C}^{M \times M}$ is a Hermitian positive definite matrix. 
When $\mathbf{v}=\mathbf{B}^{-1} \mathbf{x}$, the value of $\hbar$ is maximized. According to \eqref{SINRnew}, we let $\mathbf{v}={\mathbf{v}}_{k}[n]$, $\mathbf{x} = \sqrt{P_k} \rho_{k}[n] \widehat{\mathbf{h}}_{k}[0]$, and $\mathbf{B}=\sum_{k' \in \mathcal{K} \setminus \{k\} } P_{k'} |\rho_{k'}[n]|^{2} \widehat{\mathbf{h}}_{k'}[0] \widehat{\mathbf{h}}_{k'}^{\mathrm{H}}[0]+ \mathbf{E}$. With the help of \cite[Lemma B.4]{bjornson2017massive}, we further obtain $\mathbf{v} =\left(1+\mathbf{x}^{\mathrm{H}} \mathbf{B}^{-1} \mathbf{x}\right)\left(\mathbf{B}+\mathbf{x} \mathbf{x}^{\mathrm{H}}\right)^{-1} \mathbf{x}$. Finally, the MMSE combiner in \eqref{V_MMSE} is obtained by using the fact that the SINR in \eqref{SINRnew} remains the same if $\mathbf{v}$ is multiplied by a non-zero scalar.
\end{IEEEproof}

The MMSE combining vector \eqref{V_MMSE} minimizes the conditional mean-squared error (MSE) between ${x_k}[n]$ and the processed received signal ${\widetilde x_k[n]}$ given by  \eqref{xhat}, given the channel estimates $\{ \widehat{\mathbf{h}}_{k'}[0] \}_{k' \in\mathcal K}$. 
With MMSE combining, BS $l_k$ essentially tries to eliminate the inter-user interference by using the estimated channel information of all other users in the network. 
As discussed at the end of the previous subsection, the interference cancellation capability depends on the separability of channels between users. The more the central AoA directions are separated or the smaller the AoA spreads, the more separable the channels.  
Finally, we note that it is generally infeasible to derive \eqref{SINRnew} further into closed-form expressions. Nevertheless, the SINRs can be easily computed using the Monte Carlo method for any given ${\mathbf{v}}_{k}[n]$.

\section{Numerical Results and Discussion}
\label{sec:simulation}

In this section, the performance of the massive MIMO system is evaluated under two scenarios: the freeway scenario and the urban Manhattan grid scenario following 3GPP TR~36.885. The following metric (in bit/s/Hz/cell) is adopted for a network-level performance evaluation:
\begin{equation}
\mathrm{ASE} = \sum\limits_{k \in \mathcal{K}} {{\mathrm{SE}}_k} /L ,
\end{equation}
where $\mathrm{SE}_k$, the average achievable SE for VUE $k$ in one transmission block, is computed following Proposition \ref{prop1}. 
In the freeway scenario, the BSs are located along the freeway, 35~m away and with a fixed intersite distance (ISD), and the VUEs are all on one side of the BSs (see Fig.~A.1.3-2 in \cite{3GPP36885}). In the urban Manhattan grid scenario, the BSs are deployed in the center of each road grid and the VUEs are distributed on the edge of the cell (see Fig.~A.1.2-1 in \cite{3GPP36885}). They thus stand for two typical extreme application scenarios. Note that VUEs are placed along the centerline of the lanes following a Poisson line process.

\begin{table}[!t]
\caption{Default Simulation Parameter Settings}
\centering
\label{simupara}
\small
\begin{tabular}{l l l}
\hline
Parameter  & Freeway   &   Urban Manhattan
\\ \hline
BS number $L$      & 2        & 9                      \\
BS height          & 35 m     & 25 m                   \\
ISD                & 1732 m   & 250 m $\times$ 433 m   \\
Lane number per road        & 6        & 4  \\
Lane width         & 4 m      & 3.5 m                  \\
Sidewalk width     & /        & 3 m                    \\
Street width       & /        & 20 m                   \\
VUE speed $v$         & 33.33 m/s & 16.67 m/s             \\
VUE density\tablefootnote{Setting the minimum inter-VUE gap in each lane is $2.5 v$\cite[Table A.1.2-1]{3GPP36885}, which leading to an average number of 40 VUEs in each cell.} 
& 0.004 /m/lane  & 0.0125 /m/lane   \\
VUE moving direction $\gamma$ & See note\tablefootnote{The three-lane VUEs farther from the BS drive right, and the three-lane VUEs closer to the BS go left.} & See note\tablefootnote{VUEs travel counterclockwise along the urban Manhattan road grid.}    \\
Carrier frequency $f_c$ & \multicolumn{2}{l}{\hspace{1cm}2 GHz} \\
Symbol period $T_s$ & \multicolumn{2}{l}{\hspace{1cm}10~$\mu$s} \\
BS antenna number $M$  & \multicolumn{2}{l}{\hspace{1cm}100} \\
BS antenna spacing  $d$ & \multicolumn{2}{l}{\hspace{1cm}0.075 m ($\lambda/2$)} \\
BS antenna orientation $\alpha$ & \multicolumn{2}{l}{\hspace{1cm}0$^{\circ}$} \\
VUE transmit power $P_k$ & \multicolumn{2}{l}{\hspace{1cm}0.1 W, $\forall k\in\mathcal{K}$} \\
VUE antenna height & \multicolumn{2}{l}{\hspace{1cm}1.5 m} \\
Pilot symbol number $T$ & \multicolumn{2}{l}{\hspace{1cm}40} \\
Noise power density & \multicolumn{2}{l}{\hspace{1cm}$-$174 dBm/Hz} \\
\hline
\end{tabular}
\vspace{-4mm}
\end{table}

Following \cite{3GPP25996}, the total channel gain (in dB) is modeled as $G = -34.53-38\log_{10}(D)+X$, where $D$ is the distance between receiver and transmitter, and $X \sim \mathcal{N}\left( {0,10^2} \right)$ is the shadow fading. 
The ULA orientations for all BS are set to be the same, and the central AoD / AoA angles $\phi_{c}$ and $\theta_{c}$ are set to be the AoD and AoA of the line-of-sight (LOS) path between a BS and a VUE. In other words, the scattering clusters are considered to be centered along the LOS path, and $\phi_{c}$ and $\theta_{c}$ are  determined by the locations of the BS and VUE and $\theta_{c} = \pi + \phi_{c}$ always holds. In our simulations, the speed $v$, AoD spread $\sigma_T$, and AoA spread $\sigma_R$ of all VUEs are set to be the same for one simulation (one instance of VUE placing in the network). 
The symbol period $T_s$ is set according to $T_s = 1/B_c$, where $B_c$ stands for the coherence bandwidth given by $B_{c} =1 /\left(2 T_{d}\right)$\cite[Eq. (2.48)]{tse2005fundamentals}, and $T_{d}$ stands for the multipath delay spread. In our simulation, $T_{d} = 5$~$\mu$s is chosen, which corresponds to a maximum path-length difference of $1.5$~km, and thus $T_s = 10$~$\mu$s. The default settings for other parameters are summarized in Table \ref{simupara}. Note that wraparound is applied in simulations to mimic a large network deployment.

\begin{figure*}[!t]
	\centering
	\subfigure[MR, freeway]{\includegraphics [width= .33\linewidth]{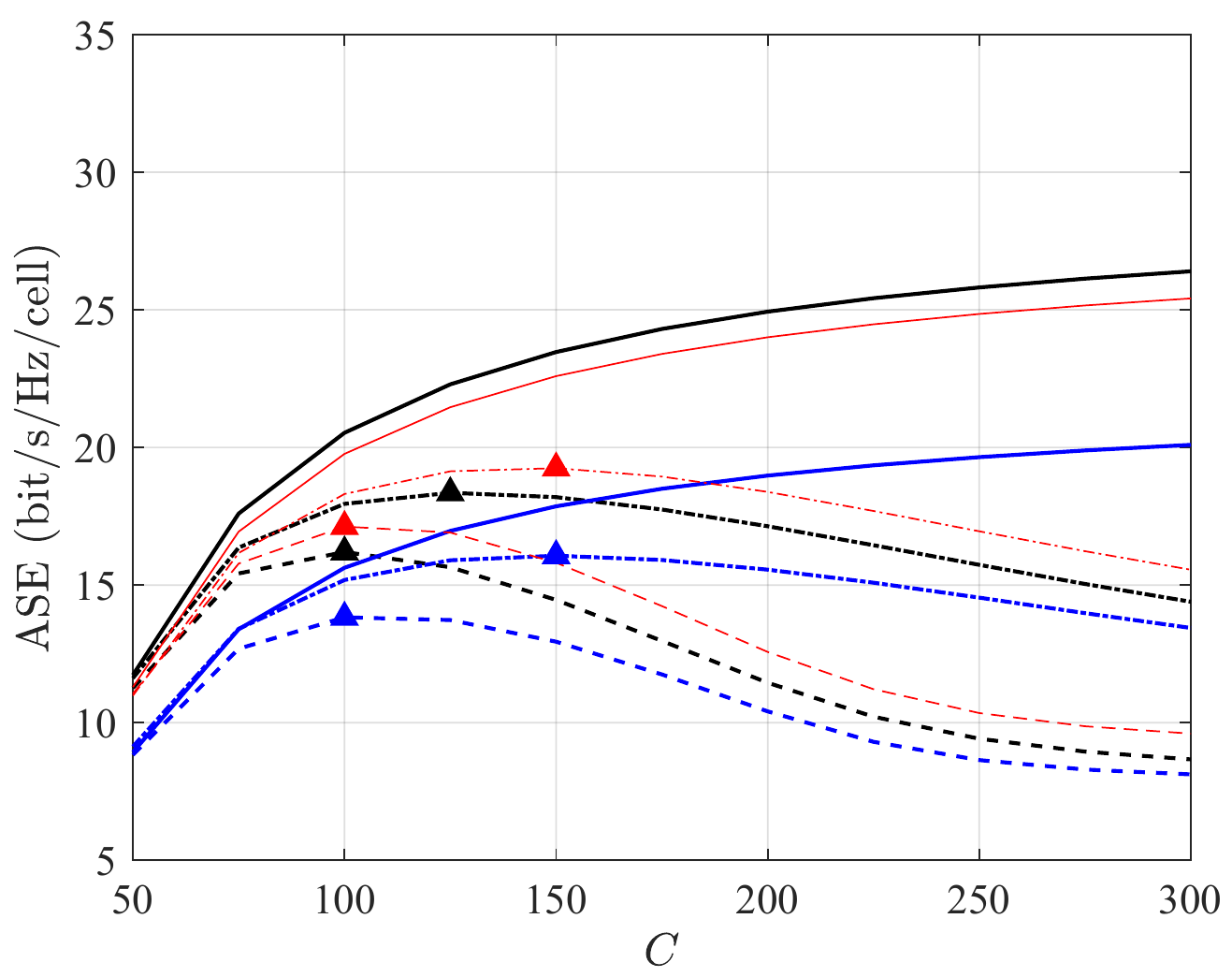} } 
	\hspace{2cm}
	\subfigure[MMSE, freeway]{\includegraphics [width= .33\linewidth]{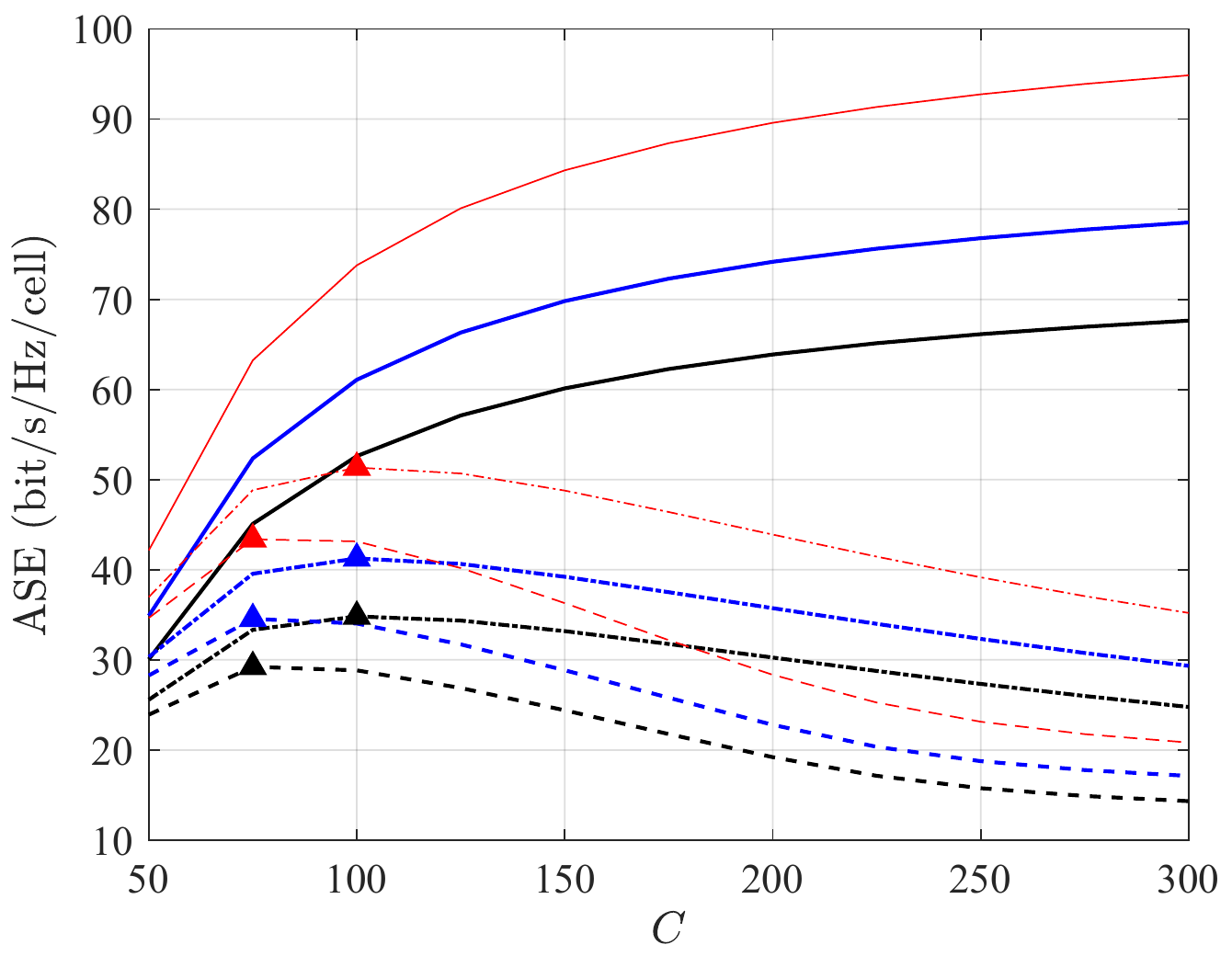} }
	\subfigure[MR, urban Manhattan]{\includegraphics [width= .33\linewidth]{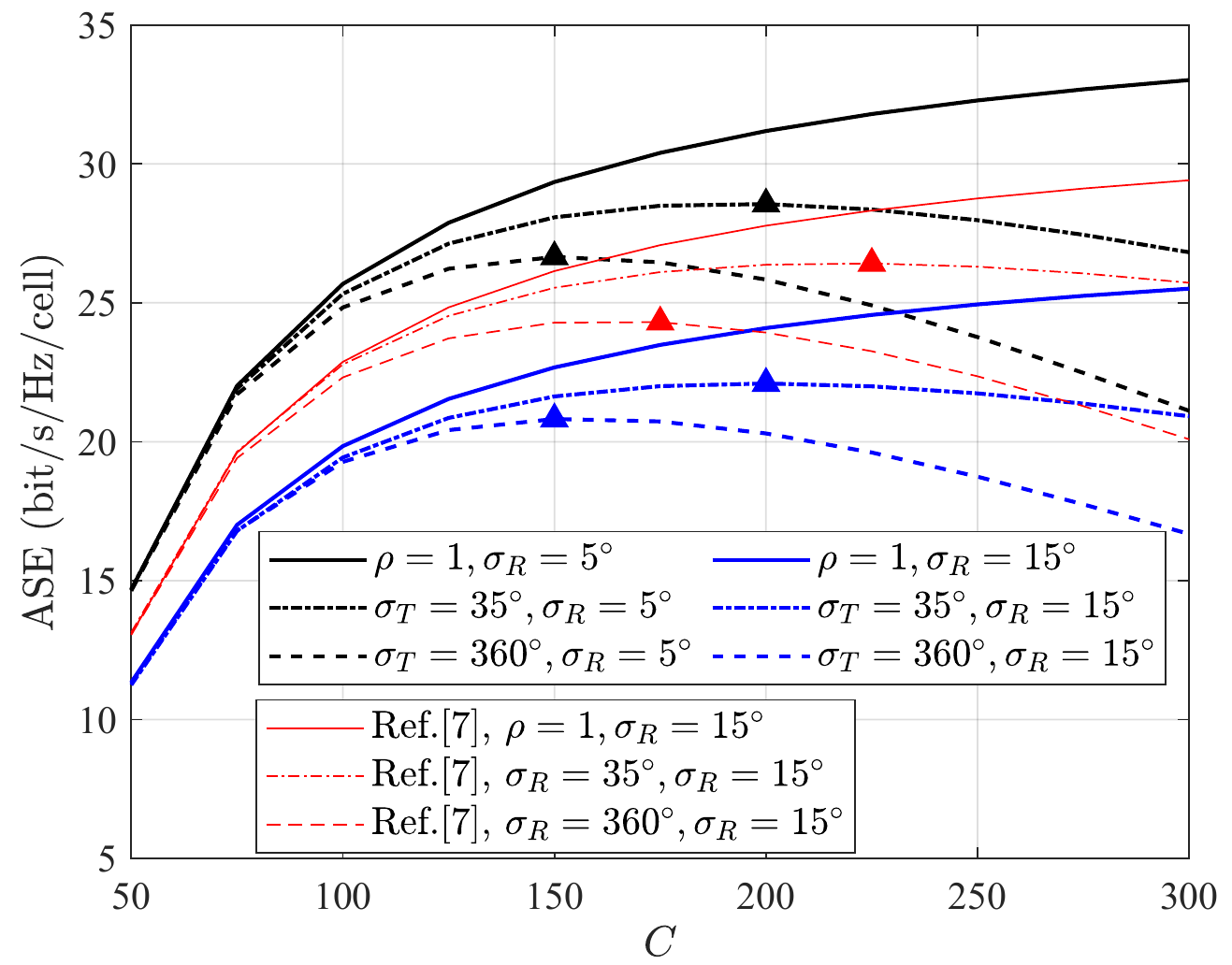} } \hspace{2cm}
	\subfigure[MMSE, urban Manhattan]{\includegraphics [width= .33\linewidth]{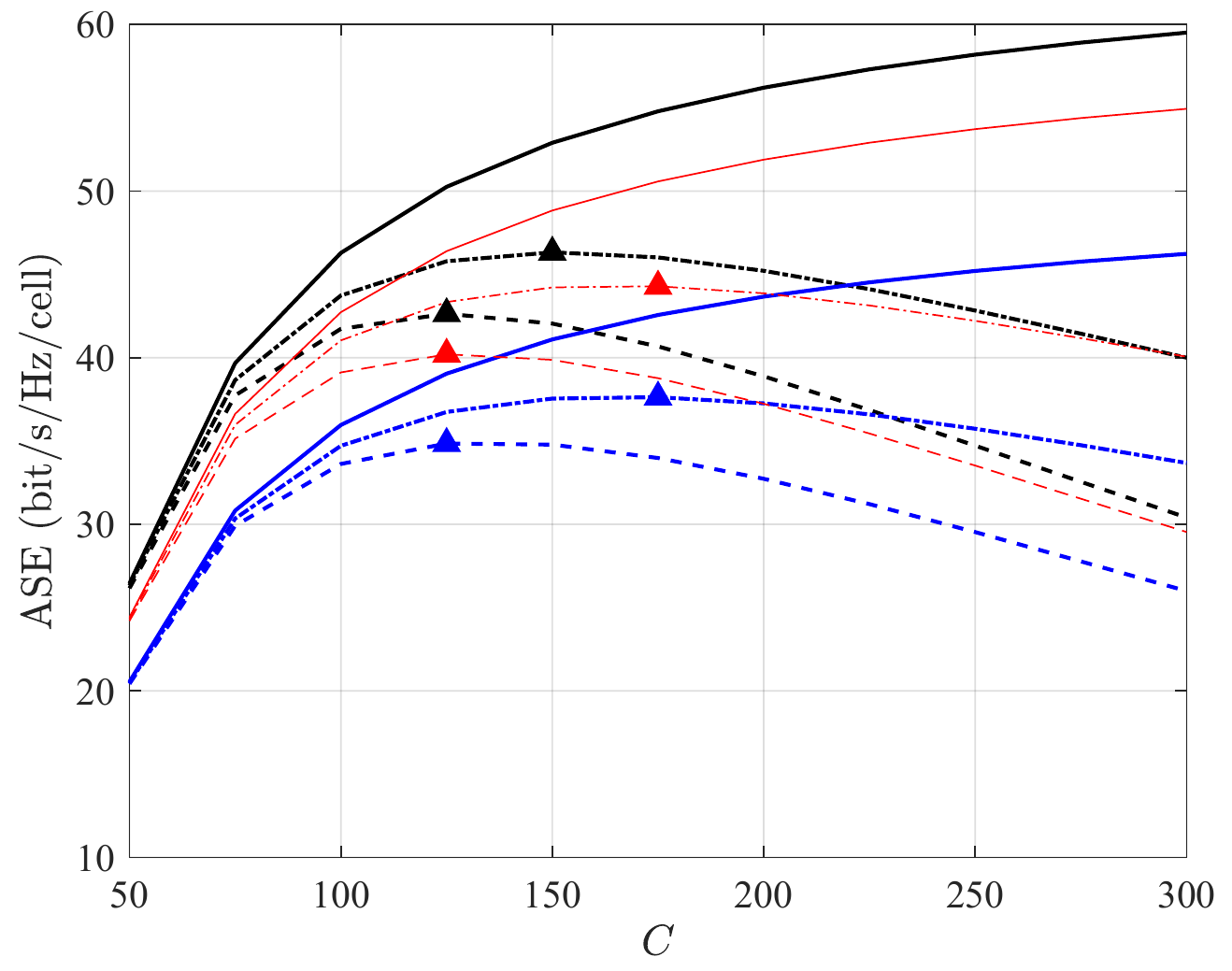} }
	\caption{The ASE performances of MR and MMSE combining with different choices of transmission block size $C$ in the different scenarios. The legend shown in (c) applies to every subplot.
	} 
	\label{fig:C_ASE} 
\end{figure*}

Fig.~\ref{fig:C_ASE} presents the ASE performance, against the transmission block size $C$, achieved by MR and MMSE combining under the two scenarios with different settings for ACF and SCF given by \eqref{ACF_SCF}, following the suggestion by \cite[Table 5.4]{WinnerII}. In particular, the BSs always see a fairly concentrated AoA distribution by setting $\kappa_R=131$ ($\sigma_R \approx 5^{\circ}$) or $\kappa_R= 14.59$ ($\sigma_R \approx 15^{\circ}$) and therefore, large values of SCF. Three different settings for the ACF are considered: 
\begin{itemize}
    \item Non-aging case by forcing $\rho=1$; 
    \item Isotropic scattering condition by letting $\kappa_T=0$ ($\sigma_T = 360^{\circ}$), which corresponds to the Jakes-Clarke model; 
    \item Non-isotropic scattering condition by setting $\kappa_T=2.68$ ($\sigma_T  \approx 35^{\circ}$). 
\end{itemize}
Note that the resulting STCCs from several combinations of the above settings have been evaluated in Section~\ref{sec:2b}.  
The ASE results achieved using the aging channel model in our preliminary study \cite{li2021impact} are also presented for comparison. Using the same notations as in Section~\ref{channelmodel}, the elements of the spatial correlation matrix adopted in \cite{li2021impact} are computed by 
\begin{equation} \label{GC_S}
\left[\mathbf{R}_0\right]_{p, q} =\frac{1}{2 \sigma_{R}} \int_{\theta_c - \sigma_R }^{\theta_c + \sigma_R} \exp ( j b \sin (\theta ) ) \,\mathrm{d} \theta ,
\end{equation} 
and the ACF is modeled by  
\begin{equation} \label{GC_T}
\rho[\tau]=\frac{I_{0}\left(\sqrt{-a^{2} + \kappa_T^{2} + 2 a j  \kappa_T  \cos(\phi_c)}\right)}{I_{0}(\kappa_T)}.
\end{equation}
We note that \eqref{GC_S} corresponds to the assumption of a uniform AoA distribution within the range of $[\theta_c - \sigma_R, \theta_c +  \sigma_R]$ as well as a fixed ULA orientation for all BSs ($\alpha = \pi/2$), and that \eqref{GC_T} corresponds to the assumption of a fixed mobility direction for all VUEs ($\gamma =0$), which can be regarded as a simplification of the ACF model given by \eqref{ACF_SCF}.

It is trivial to infer that if the channel is modeled as non-aging ($\rho=1$), the ASE curves (the solid lines in Fig.~\ref{fig:C_ASE}) will increase continuously with $C$. When the aging effect is captured by the channel model, one can expect that there exists an optimal transmission block size $C_{\mathrm{opt}}$ that maximizes ASE (as marked by the solid triangles on the dashed and dotted ASE curves). 
The reason is also clear: Our knowledge of the actual channel coefficient from channel estimation decreases with time (see the channel model given by \eqref{realchannel}--\eqref{E_kl}), resulting in a decrease in instantaneous SINR in \eqref{SINRnew}. When the channel ages to a certain extent, it becomes worthwhile to perform channel training again at the cost of the $T$ symbols. 

Furthermore, Fig.~\ref{fig:C_ASE} shows that the achieved ASE performance and the obtained $C_{\mathrm{opt}}$ depend greatly on the AoA/AoD distribution, the application scenario (which determines VUEs' mobility and the relative spatial distribution relation of BSs and VUEs), and the combining strategy under consideration. A smaller AoD spread can in general better resist the channel aging effect and lead to larger ASE. The isotropic scattering assumption results in an overly pessimistic performance prediction and suggests small values of $C_{\mathrm{opt}}$, which may lead to a less efficient system design, as we pointed out in \cite{li2021impact}. 
On the other hand, the impact of the AoA distribution condition, and hence the spatial correlation of the massive MIMO channel, on the ASE performance is more complicated. By comparing the black-colored curves with the blue ones, we can see that a more concentrated AoA distribution (small $\sigma_R$) leads to better ASE performance to all the subfigures but subfigure (b): the freeway scenario with MMSE combing. We have specifically discussed this issue in Section \ref{sect:scf_impact}. A very concentrated AoA distribution ($\sigma_R \approx 5^{\circ}$) may help to resist the pilot contamination effect during the channel training phase (It was also found in \cite{Channel_Estimation6415397} and \cite{9276421} that strong spatial correlation helps to improve multi-user system performance), but apparently fail in the freeway scenario because the central AoA directions of all the VUE-BS links are close to each other. This eventually deteriorates the performance of MMSE combining, which relies heavily on good channel estimation. 
We also remark that the degradation of the ASE performance due to channel aging is also the most severe in this scenario, owing largely to the high mobility of the VUEs. 

Finally, we note that a clear gap between the ASE curved obtained using the aging channel model in \cite{li2021impact} and the model developed in this paper can be observed, although similar  trends are shared. This emphasizes again the importance of employing a channel model that correctly reflects the spatial and temporal correlation of the channel coefficients.

Evaluation of $C_\mathrm{opt}$, the optimal transmission block size, plays a vital role in the design and optimization of a massive MIMO system. In \cite{li2021impact}, we showed that empirically, $C_{\mathrm{opt}}$ can be closely modeled as a linear equation of {${\kappa_T}$} with scenario-dependent coefficients. Above we have shown that the impact of spatial correlation caused by the AoA distribution is also profound. The channel model developed in Section~\ref{sec:system_model} allows us to assess the impact of AoA and AoD jointly and also refine the empirical model of $C_{\mathrm{opt}}$.

According to the study by WINNER II \cite[Table 5.4]{WinnerII} and 3GPP \cite[Table 5.1]{3GPP25996}, we make the value of $\kappa_T$ and $\kappa_R$ to change such that correspondingly, $\sigma_T$ and $\sigma_R$ vary within the range of $[5^{\circ}, \ 50^{\circ}]$ (which is suitable for outdoor non-line-of-sight (NLOS) scenarios). This is applied to both the freeway and the urban Manhattan scenarios. The speed of the VUEs changes within the range of $[19.44, \ 38.89]$ m/s ($[70, \ 140]$ km/h) for the freeway scenario and $[8.33, \ 33.33]$ m/s ($[30, \ 120]$ km/h) for the urban Manhattan scenario. For each scenario and each particular choice of $(\sigma_T, \sigma_R, v)$, the ASE is calculated over a wide enough range of $C$, and $C_{\mathrm{opt}}$ is identified as the value of $C$ that maximizes the resulting ASE. Through nonlinear regression, it is found that $C_{\mathrm{opt}}$ can be closely approximated by linear equations of $v$, $\sqrt{\sigma_T}$, and $\sqrt{\sigma_R}$ ({$v$ is in [m/s], and $\sigma_T$ and $\sigma_R$ is in [degree]}). These obtained empirical models, denoted by $C_{\mathrm{opt}}^*$, are summarized in Table~\ref{C_opt_eq}. To note that coefficients of the model are significantly distinct for different scenarios and different combining strategies.

\begin{table*}[!t] 
\caption{Empirical Model $C_{\mathrm{opt}}^*$ Under Freeway and Urban Manhattan Scenarios with MR/MMSE Combining, {where $v$ is in [m/s], $\sigma_T$ and $\sigma_R$ is in [degree]}. } 
\vspace{-0.2cm}
\label{C_opt_eq}
\centering
\begin{tabular}{|c|c|c|}
\hline
 & $C_{\mathrm{opt,MR}}^*(v, \sigma_T, \sigma_R)$  & $C_{\mathrm{opt,MMSE}}^*(v, \sigma_T, \sigma_R)$ \\
\hline \hline
Freeway & \hspace{1em} $568.91-5.51v-53.92\sqrt{\sigma_T}+4.04\sqrt{\sigma_R}$ \hspace{1em} &   \hspace{1em} $298.24-1.35v-20.14\sqrt{\sigma_T}-7.72\sqrt{\sigma_R}$ \hspace{1em} \\ 
\hline
Urban Manhattan & \hspace{1em}$613.79-6.03v-50.18\sqrt{\sigma_T}+0.63\sqrt{\sigma_R}$ \hspace{1em}   & \hspace{1em}$522.58-5.42v-46.94\sqrt{\sigma_T}+7.61\sqrt{\sigma_R}$ \hspace{1em}  \\
\hline
\end{tabular}  
\vspace{-4mm}
\end{table*}

\begin{table}[!t]
	\caption{${\bar R^2}$ and $\mathrm{NRMSE}$ Values.}
	\vspace{-0.2cm}
	\centering
	\begin{tabular}{|c|c|c|c|c|} 
		\hline
		{} & \multicolumn{2}{c|}{${\bar R^2}$} & \multicolumn{2}{c|}{$\mathrm{NRMSE}$}\\    \hline\hline 
        Combining Method  &      MR           & MMSE     & MR          & MMSE \\
        \hline
        Freeway                  & $0.92$       & $0.89$     & $0.08$        & $0.06$     \\ 
        \hline
        Urban Manhattan          & $0.92$       & $0.90$     & $0.09$       & $0.09$     \\
        \hline
	\end{tabular}
\vspace{-4mm}
\label{table:fitness}
\end{table}

To prove the closeness of the fit, Table~\ref{table:fitness} presents the results of the coefficient of determination metric (${\bar R^2}$) and the normalized root mean square error metric (${\mathrm{NRMSE}}$), defined as follows:
\begin{align}
    &\bar{R}^{2} =1-\frac{\sum\nolimits_{i \in \mathcal{S}}\left|C_{\mathrm{opt}}(i)-C_{\mathrm{opt}}^{*}(i)\right|^{2}}{\sum\nolimits_{i \in \mathcal{S}}\left|C_{\mathrm{opt}}(i)-\frac{1}{|\mathcal{S}|} \sum\nolimits_{i \in \mathcal{S}} C_{\mathrm{opt}}(i)\right|^{2}}, \\
    &\mathrm{NRMSE} = \frac{1}{N_r} \sqrt{\frac{\sum\nolimits_{i \in \mathcal{S}}\left|C_{\mathrm{opt}}(i)-C_{\mathrm{opt}}^{*}(i)\right|^{2}}{|\mathcal{S}|}},
\end{align}
where $\mathcal{S}$ stands for the set of all ($\sigma_R, \sigma_T, v$) samples adopted in the simulation and $|\mathcal{S}|$ is its cardinality, and $N_r$ used for normalization is given by the range of $C_{\mathrm{opt}}$, namely, $N_r=\max(C_{\mathrm{opt}}) - \min(C_{\mathrm{opt}})$. 
Note that ${\bar R^2} \in [0,1]$ and $\mathrm{NRMSE} \ge 0$, and that a large $\bar R^2$ or a small $\mathrm{NRMSE}$ implies a good fit.

The models given in Table~\ref{C_opt_eq} first show that although $C_\mathrm{opt}$ is affected by $v$, $\sigma_T$ and $\sigma_R$ together, the impact of $\sigma_T$ is considerably more significant than the other two. 
Second, generally speaking, MR combining allows larger $C_\mathrm{opt}$ than MMSE combining. The gap is especially large under the freeway scenario. This shows that MR combining has a much better resilience to channel aging than MMSE combining. 
Third, it can be seen that for the freeway scenario with MMSE combining, the value of $C_\mathrm{opt}$ is considerably smaller than the rest of the settings, and it is the only case that the coefficient in front of 
$\sqrt{\sigma_R}$ is negative in the fitted model. This reminds us again of the extreme spatial distribution of VUEs and the complicated effects of spatial correlation on different aspects of the system, and shows again that in this setting, a smaller AoA spread (and thus larger spatial correlation) is beneficial in the sense of a large $C_\mathrm{opt}$. Nevertheless, the claimed performance by MMSE comes at a higher cost (more frequent channel estimation) in this case.

\begin{figure*}[!t]
	\centering
	\subfigure[$\sigma_R = 15^{\circ}$ ($\kappa_R \approx 14.59$)]{\includegraphics [width= .30\linewidth]{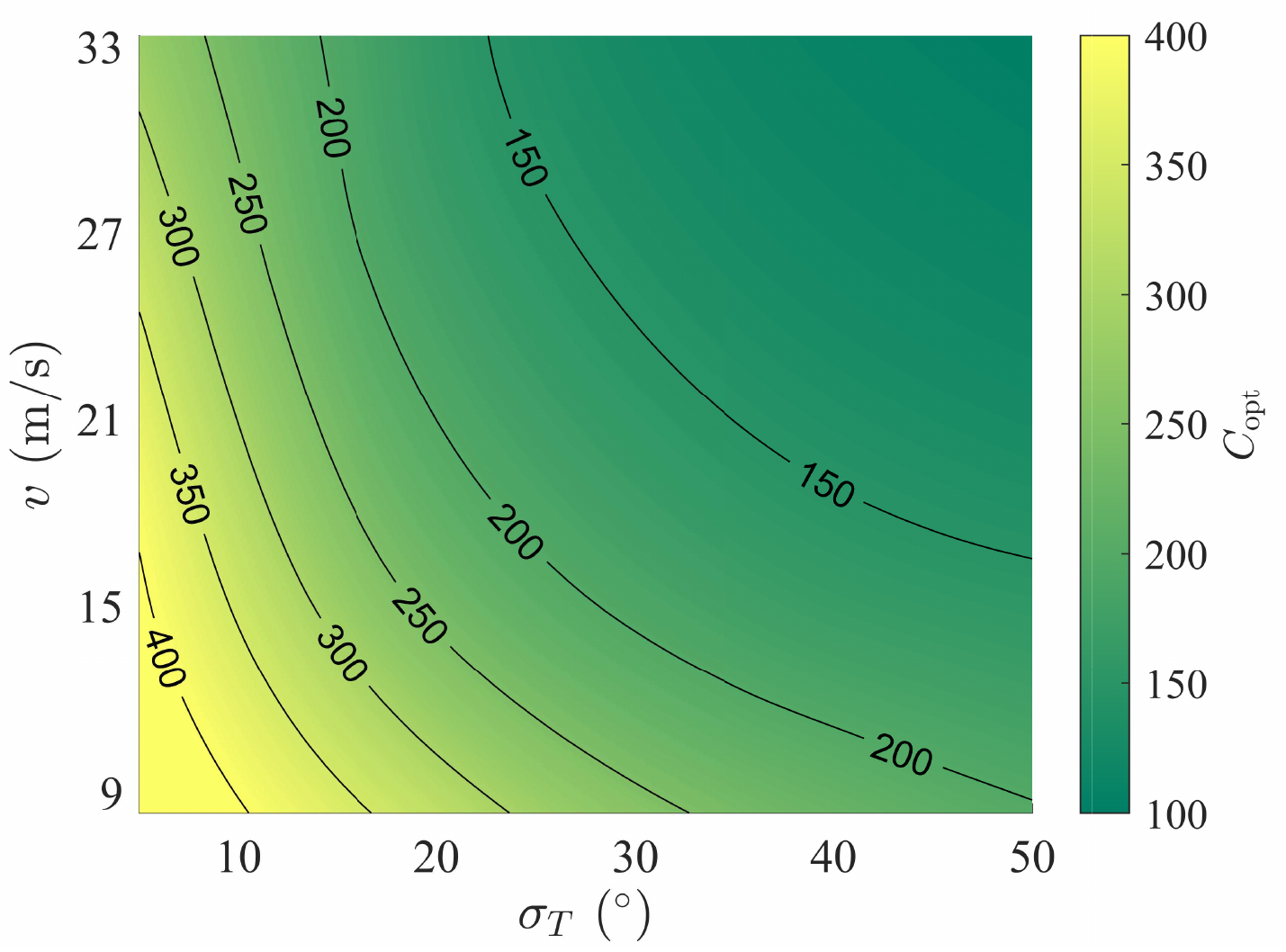} } 
	\subfigure[$\sigma_T = 35^{\circ}$ ($\kappa_T \approx 2.68$)]{\includegraphics [width= .30\linewidth]{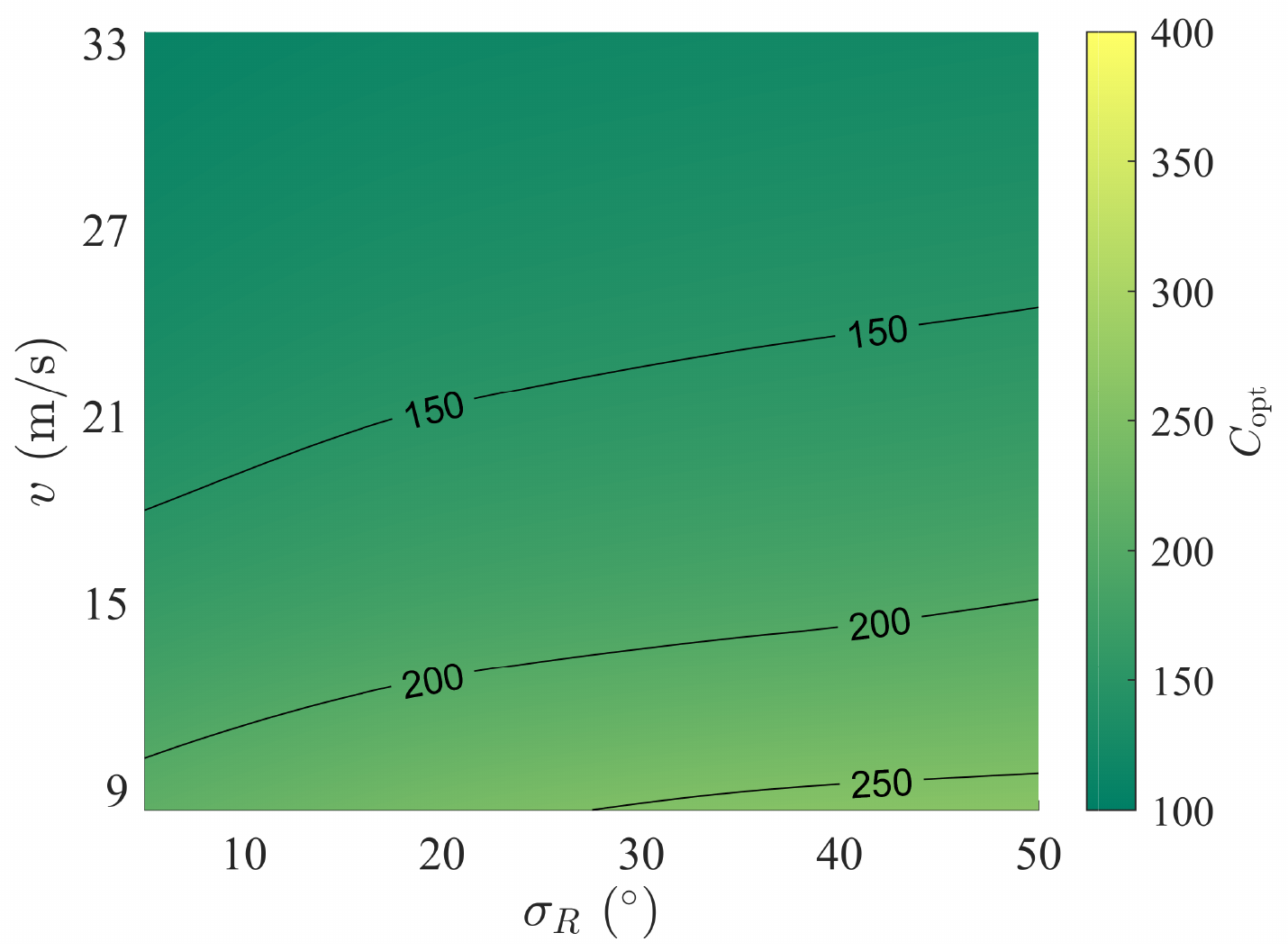} }
	\subfigure[$v = 16.67$ m/s ($60$ km/h)]{\includegraphics [width= .30\linewidth]{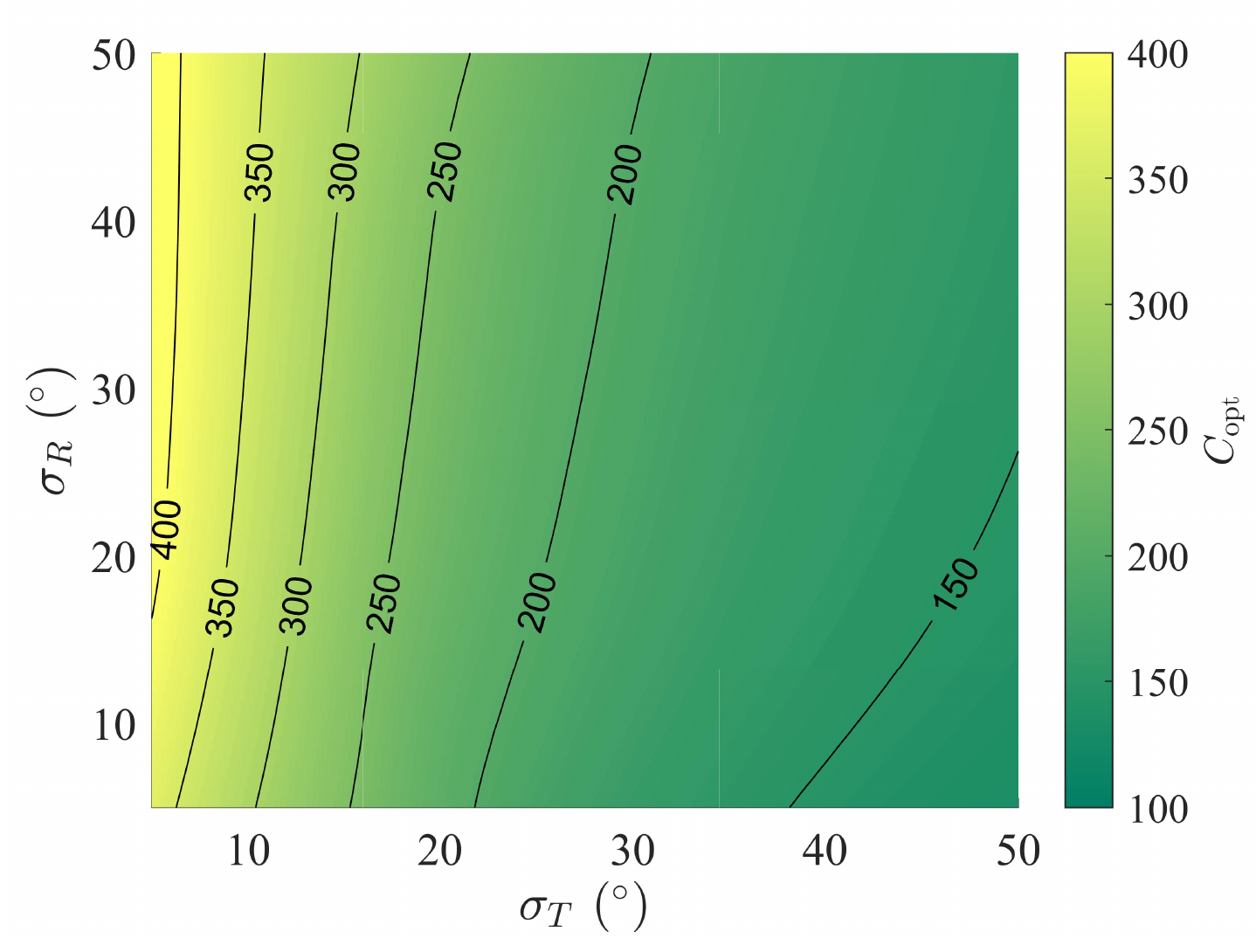} }
	\caption{The results of $C_\mathrm{opt}$ for the urban Manhattan scenario with MMSE combining, for some fixed $\sigma_R$, $\sigma_T$, $v$ values. 
	} 
	\label{fig:C_opt} 
\end{figure*}

As a demonstration, we show in Fig.~\ref{fig:C_opt} the results of the obtained $C_{\mathrm{opt}}$ under the urban Manhattan scenario with MMSE combining. The three color-map type subfigures illustrate how $C_{\mathrm{opt}}$ changes with two out of the three parameters, while having $\sigma_R = 15^{\circ}$ ($\kappa_R \approx 14.59$), $\sigma_T = 35^{\circ}$ ($\kappa_T \approx 2.68$), and $v = 16.67$ m/s ($60$ km/h) fixed, respectively. It can be shown that 
using the empirical models of $C_{\mathrm{opt}}^*$ in Table \ref{C_opt_eq}, the obtained curves achieving the same ASE values are very close to the contour lines shown in the figure.

\begin{figure*}[!t]
	\centering
	\subfigure[$\sigma_R = 15^{\circ}$ ($\kappa_R \approx 2.68$)]{\includegraphics [width= .30\linewidth]{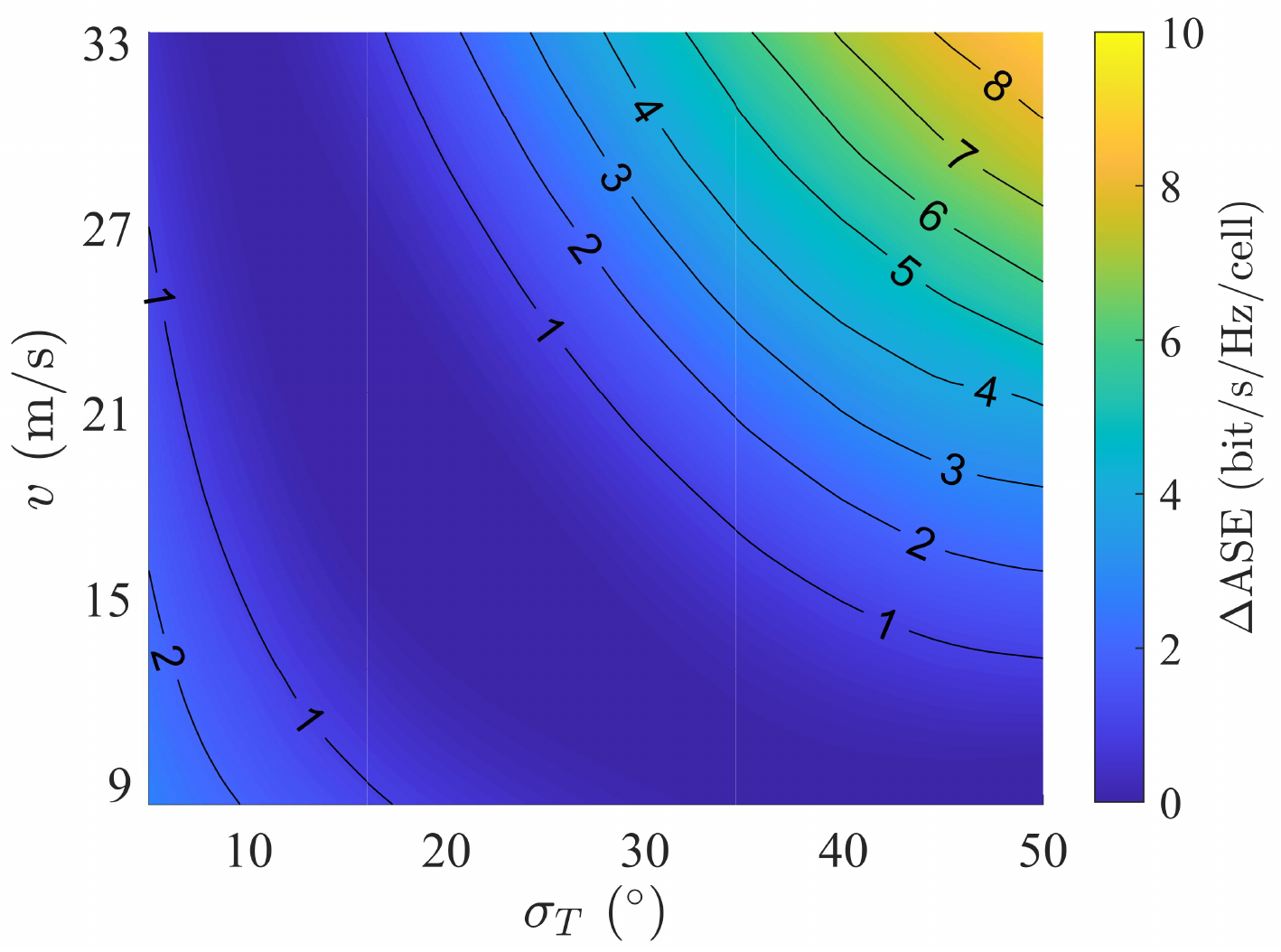} } 
	\subfigure[$\sigma_T = 35^{\circ}$ ($\kappa_T \approx 14.59$)]{\includegraphics [width= .30\linewidth]{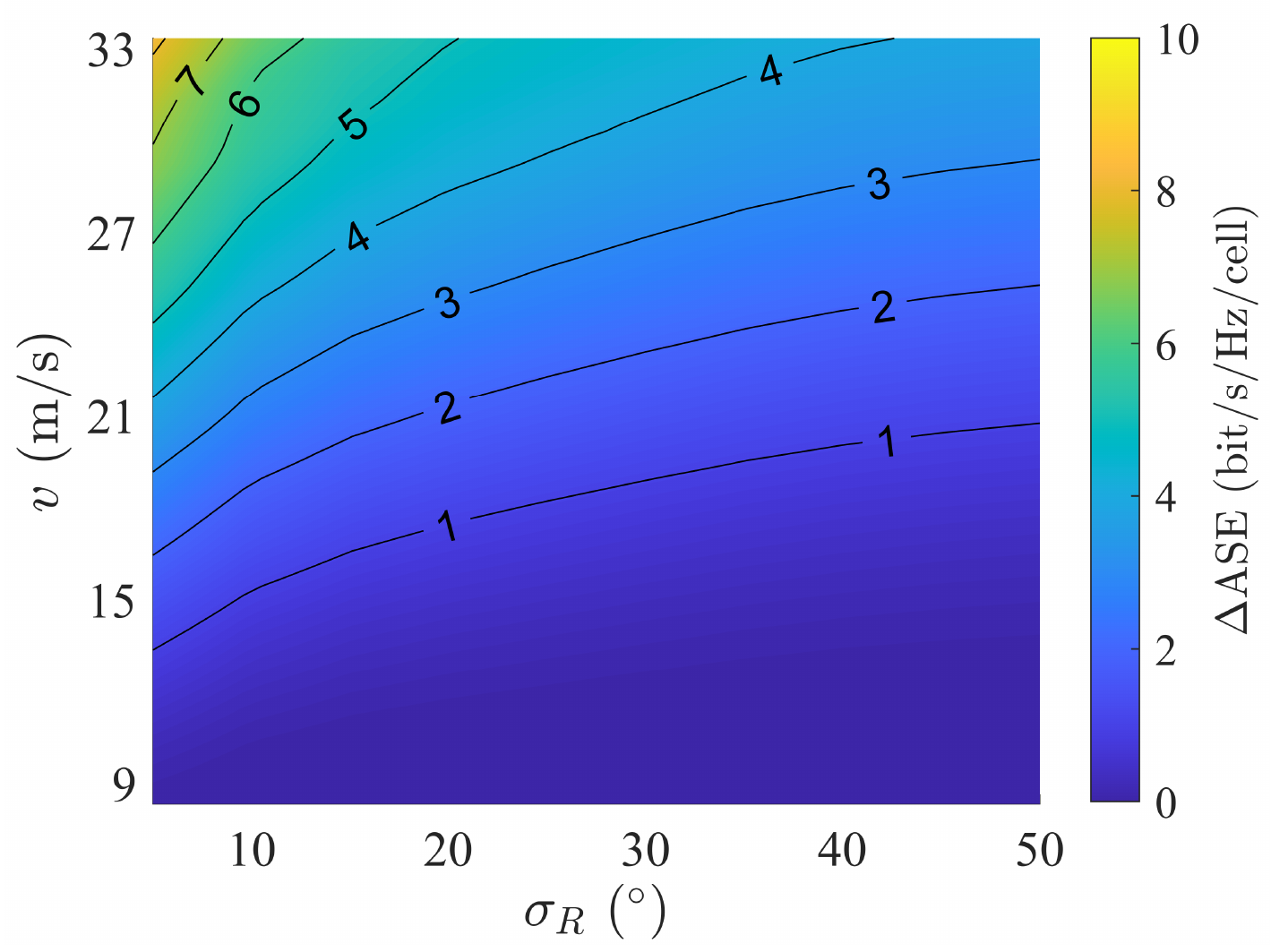} }
	\subfigure[$v = 16.67$ m/s ($60$ km/h)]{\includegraphics [width= .30\linewidth]{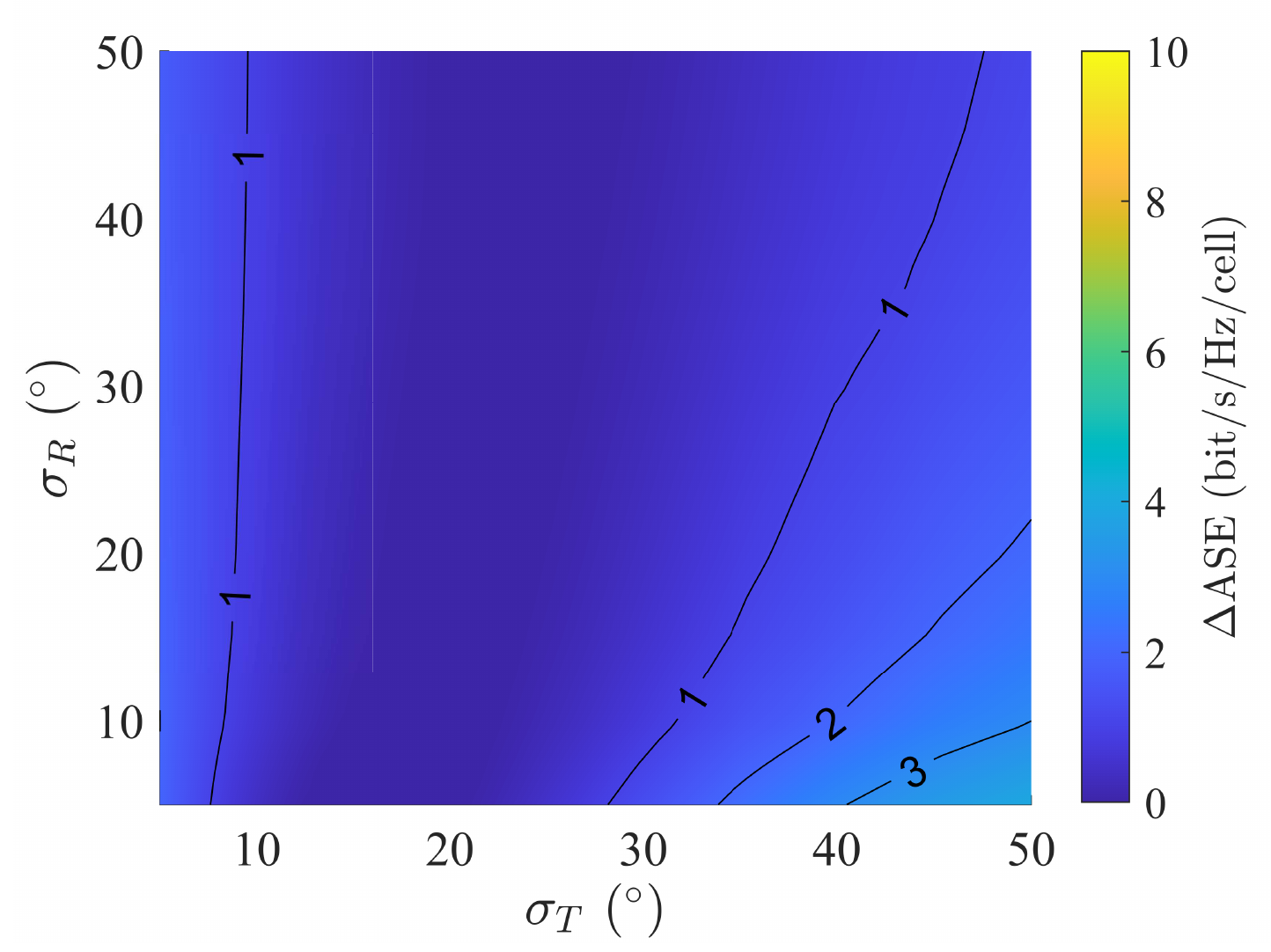} }
	\caption{The results of $\Delta \mathrm{ASE}$ for the  urban Manhattan scenarios with MMSE combining, for some fixed $\sigma_R$, $\sigma_T$, $v$ values.
	} 
	\label{fig:delta_ASE} 
\end{figure*}

Finally, we demonstrate the performance gain in ASE brought about by adopting $C_{\mathrm {opt}}^*$ (following models in Table~\ref{C_opt_eq}) over the coherence-time-based block size design (which is currently adopted by the 4G/5G cellular networks \cite{3GPP38211,marzetta2016fundamentals}), denoted by $C_v$ in what follows.  $C_v$ is determined only by the Doppler spread, caused by the mobility of the users. To be specific, the duration of a transmission block, i.e., $C T_s$, is set to be equal to the coherence time, given by $T_c = \lambda /\left( {4 v} \right)$ \cite[Eq. (2.45)]{tse2005fundamentals}. Variations in the phase and amplitude of multipath components due to users' mobility are considered negligible in $T_c$, and therefore the channel is considered to be unchanged. Note that the default speed settings given in Table~\ref{simupara} lead to $C_v =113$ for the freeway scenario and $C_v=225$ for the urban Manhattan scenario ($T_s = 10$~$\mu$s in our setup). 
The ASE difference achieved by the two designs, namely, 
\begin{equation}
    \Delta \mathrm{ASE} = \mathrm{ASE}\left({C}_{\mathrm{opt}}^* \right) - \mathrm{ASE}\left( C_v \right)
\end{equation}
is adopted as the metric. Performance gain is observed for all scenarios and combining methods. As an example, the results are shown in Fig.~\ref{fig:delta_ASE} for the urban Manhattan scenario with MMSE combining, where the significant gain from $C_{\mathrm {opt}}^*$ can be clearly seen. 

\section{Conclusion}
\label{sec:conclusion}

In this paper, a novel aging channel model has been developed for the wireless channel between a VUE and a BS with ULA in a multi-cell massive MIMO network based on the single cluster scattering assumption. In particular, two independent von Mises distributions have been adopted to describe the angular distributions of the scatterers seen by BS and VUE.  
Based on the aging channel model, expressions for the user SE have been derived for both MR and MMSE combining for UL transmission and the system-level ASE performance of the massive MIMO network has been evaluated numerically under the freeway and urban Manhattan grid scenarios with 3GPP-recommended settings. This numerical study has also lead to insightful and easy-to-use empirical optimal transmission block size models --- linear equations of VUE moving speed and square roots of AoD and AoA spread, with scenario-dependent coefficients. The substantial performance gain brought by these empirical model has been demonstrated. 

The developed aging channel model enables the study the joint effect of spatial and temporal (aging) correlations of the wireless channel on the performance of the massive MIMO network and captures the impact of the spatial distribution of BS, VUE, and scatterers, as well as the BS antenna array setting and the VUE mobility. More importantly, heterogeneous settings are allowed for numerical study, making the model applicable to a wide range of scenarios. Following the same approach employed in our numerical study, empirical models for the optimal design of the transmission block can also be obtained for different scenarios and parameter settings, and hopefully provide some valuable guidance to the optimization of system operating efficiency. 

\bibliographystyle{IEEEtran}
\bibliography{bibFile/mylib}

\end{document}